\documentclass[11pt,twoside]{article}
 \let\openright=\cleardoublepage
\usepackage{ae,aecompl}
\usepackage{helvet}
\usepackage[T1]{fontenc}
\usepackage{geometry}
\usepackage{color}
\usepackage{float}
\usepackage{textcomp}
\usepackage{amsbsy}
\usepackage{amstext}
\usepackage{amssymb}
\usepackage{graphicx}
\usepackage{setspace}
\usepackage{esint}
\usepackage{url}
\usepackage{enumitem}
\usepackage[authoryear]{natbib}
\usepackage[cp1250]{inputenc}
\usepackage{pdfpages}
\usepackage[nottoc]{tocbibind}

\usepackage{hyperref} 
\hypersetup{pdfpagemode=UseNone}
\hypersetup{pdfstartview={XYZ null null 1.00}}
\hypersetup{pdftitle=Galaxy interactions: dark matter vs. Modified Newtonian dynamics (MOND)}
\hypersetup{pdfauthor=Michal B{\' i}lek}

\newcommand{\apjs}{ApJ}
\newcommand{\apjl}{ApJ}

\newcommand{\aap}{A\&A}

\renewcommand{\d}{\mathrm{d}}
\newcommand{\sers}{S{\' e}rsic }
\newcommand{\equ}[1]{Eq.~\ref{eq:#1}}
\newcommand{\fig}[1]{Fig.~\ref{fig:#1}}
\newcommand{\tab}[1]{Tab.~\ref{tab:#1}}
\newcommand{\sect}[1]{Sect.~\ref{sec:#1}}
\newcommand{\app}[1]{Appendix~\ref{ap:#1}}

\newcommand{\atlas}{ATLAS$^\mathrm{3D}\,$} 
 
\newcommand{\ngc}{NGC\,3923}
\newcommand{\arcsec}{$^{\prime\prime}$}

\begin{document}


\pagestyle{empty}
\begin{center}

\large

Charles University in Prague

\medskip

Faculty of Mathematics and Physics

\vfill

{\bf\Large DOCTORAL THESIS}

\vfill

\centerline{\mbox{\includegraphics[width=60mm]{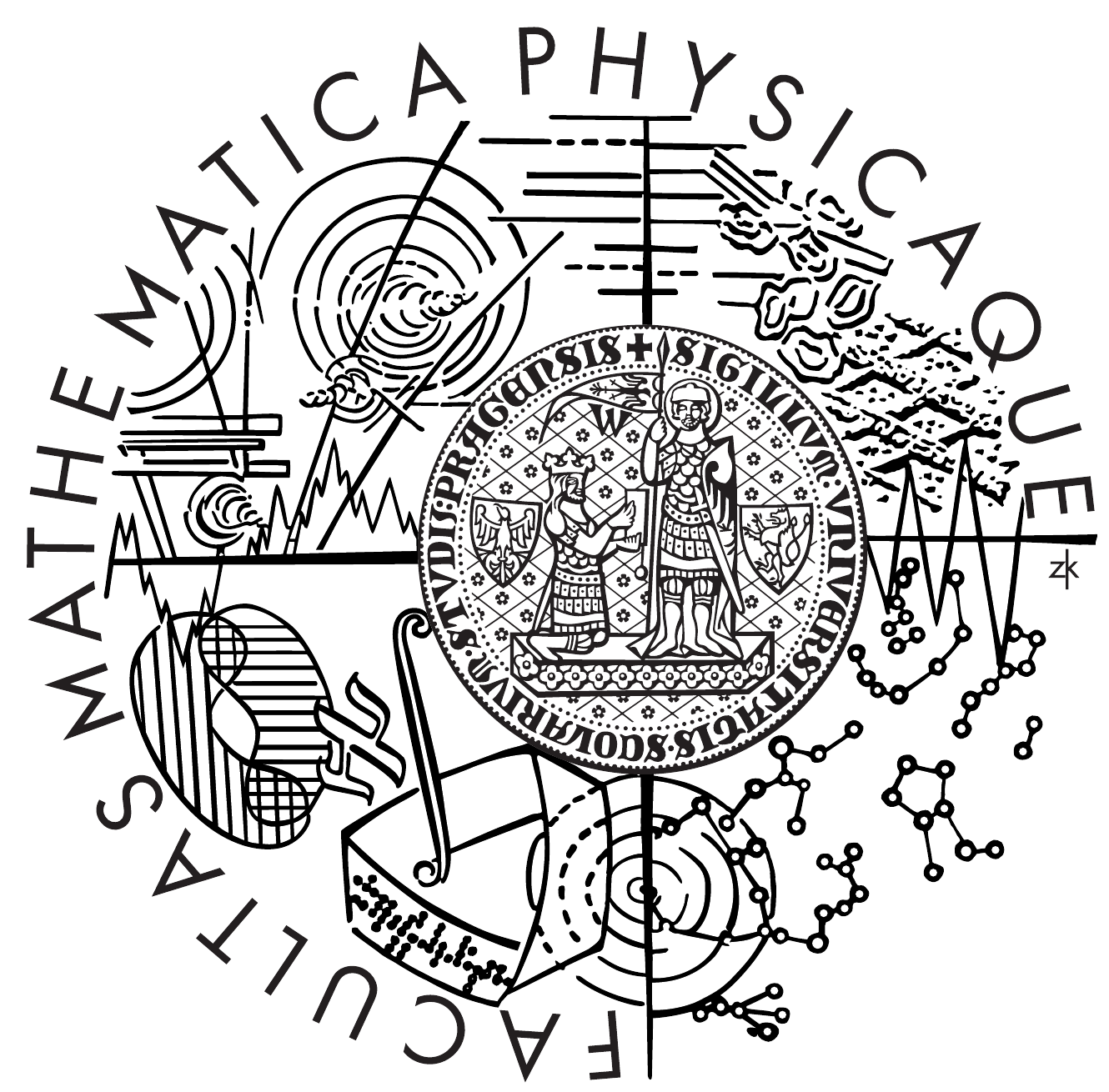}}}

\vfill
\vspace{5mm}

{\LARGE Michal B{\' i}lek}

\vspace{15mm}

{\LARGE\bfseries Galaxy interactions: dark matter vs. Modified Newtonian dynamics (MOND)}

\vfill

Astronomical Institute, ASCR

\vfill

\begin{tabular}{rl}

Supervisor of the doctoral thesis: & Bruno Jungwiert \\
\noalign{\vspace{2mm}}
Study programme: & Physics \\
\noalign{\vspace{2mm}}
Specialization: & Theoretical Physics, Astronomy and Astrophysics \\
\end{tabular}

\vfill

Prague 2015

\end{center}

\newpage


\vglue 0pt plus 1fill

\noindent
I declare that I carried out this doctoral thesis independently, and only with the cited
sources, literature and other professional sources.

\medskip\noindent
I understand that my work relates to the rights and obligations under the Act No.
121/2000 Coll., the Copyright Act, as amended, in particular the fact that the Charles
University in Prague has the right to conclude a~license agreement on the use of this
work as a~school work pursuant to Section 60 paragraph 1 of the Copyright Act.

\vspace{10mm}

\hbox{\hbox to 0.5\hsize{%
In \underline{\textit{Prague}}\hspace{1em} date \underline{\textit{June 19, 2015}}

\hss}\hbox to 0.5\hsize{%
signature of the author
\hss}}

\vspace{20mm}
\newpage


\vbox to 0.5\vsize{
\setlength\parindent{0mm}
\setlength\parskip{5mm}

Název práce:
Galaxy interactions: dark matter vs. Modified Newtonian dynamics (MOND)

Autor:
Mgr. Michal Bílek

Katedra:  
Astronomický ústav AV ÈR, v.v.i.

Vedoucí disertaèní práce:
RNDr. Bruno Jungwiert, Ph.D., Astronomický ústav AV ÈR, v.v.i.

Abstrakt:
MOND je observaènì zjistìné pravidlo pro pøedvídání zrychlení hvìzd a~galaxií na základì rozloení pozorovatelné hmoty. Moná je to dùsledek nového zákona fyziky. Shrnuji teoretické stránky MONDu, jeho výsledky a~problémy. MOND byl doposud testován hlavnì v~diskových galaxiích. Jeho testy v~eliptických galaxiích jsou vzácné, protoe projevy MONDu jsou u nich malé v~oblastech pozorovatelných obvyklými metodami. V~práci vysvìtluji medtody a~nápady, které jsem vytvoøil pro testování MONDu v~eliptických galaxiích pomocí hvìzdných slupek. Slupky nám navíc poprvé umoòují testovat MOND pro hvìzdy na radiáních drahách. Slupky jsou výsledkem galaktických interakcí. Vysvìtluji mechanismy jejich tvorby a~shrnuji výsledky z pozorování a~simulací slupek.

Klíèová slova:
Astrofyzika, Galaxie, Gravitace, Dynamika hvìzd, MOND

\vss}\nobreak\vbox to 0.49\vsize{
\setlength\parindent{0mm}
\setlength\parskip{5mm}

Title:
Galaxy interactions: dark matter vs. Modified Newtonian dynamics (MOND)

Author:
Mgr. Michal B{\' i}lek

Department:
Astronomical Institute, ASCR

Supervisor:
RNDr. Bruno Jungwiert, Ph.D., Astronomical Institute, ASCR

Abstract:
MOND is an observational rule for predicting the acceleration of stars and galaxies from the distribution of the visible matter. It possibly stems from a~new law of physics. I list the theoretical aspects of MOND, its achievements and problems. MOND has been tested mainly in disc galaxies so far. Its tests in elliptical galaxies are rare because the MOND effects are small for them in the parts observable by the conventional methods. In the thesis, I explain the methods and ideas I developed for testing MOND in the ellipticals using stellar shells. Moreover, the shells enable us to test MOND for stars in radial orbits for the first time. The shells are results of galactic interactions. I discuss the shell formation mechanisms and summarize the findings from shell observations and simulations.


\vss}

\newpage


\openright
\pagestyle{plain}
\setcounter{page}{1}
\tableofcontents
\newpage
\section{MOND}
\subsection{Motivation for MOND}\label{sec:mond}
As we know, physics encountered the missing mass problem (MMP). Gravitational attraction between cosmic objects is evidently higher than we expect in General Relativity or Newtonian dynamics from the distribution of the observable matter.  There are two ways to solve the MMP. The first is to postulate the existence of dark matter (DM) which demonstrates its existence only by the gravitational influence. The amount of DM in the Universe is then much larger than that of the observable matter. There is evidence that DM cannot by made of any known kind of particles. These particles are not expected by the Standard model of particle physics. Dark matter has great success in explaining the properties of large cosmological structures (cosmic microwave background -- \citealp{planck}, baryon acoustic oscillations -- \citealp{bao}, the relation between the SN Ia magnitudes and redshifts -- \citealp{snia}, Big Bang nucleosynthesis -- \citealp{nucleo}, but see \citealp{mcgaugh08}). The cosmological parameters come out in accordance with each other from these observations.  The DMF also encounters problems (missing satellites p. -- \citealp{klypin99}, core-cusp p. -- \citealp{deblok10}, too big to fail p. -- \citealp{boylan11}, disk of satellites p. -- \citealp{ibata13}, missing baryons p. -- \citealp{mcgaugh08}, missing DM p. -- \citealp{karachentsev12}, the formation of bulgeless galaxies -- \citealp{mayer08}, too massive galaxy clusters at high redshifts -- \citealp{gonzalez12}, too high collision velocities of some galaxy clusters  -- \citealp{lee10}). There are ongoing efforts to detect the DM particles in laboratory and space. We will call General Relativity with DM added as the dark matter framework (DMF) hereafter.

The other way to solve the MMP is to assume that we are able to detect most of the matter in the Universe. Then we have to modify some of the standard laws of physics. These laws were derived from laboratory experiments and the Solar system observations. But on the galactic scales, many quantities (acceleration, angular momentum, mean density, \ldots) take values different from those in these experiments by many orders of magnitude. By applying the standard laws on galaxies, we make big extrapolations of what has been tested reliably.

\begin{figure}
	\centering\includegraphics[width=0.9\textwidth]{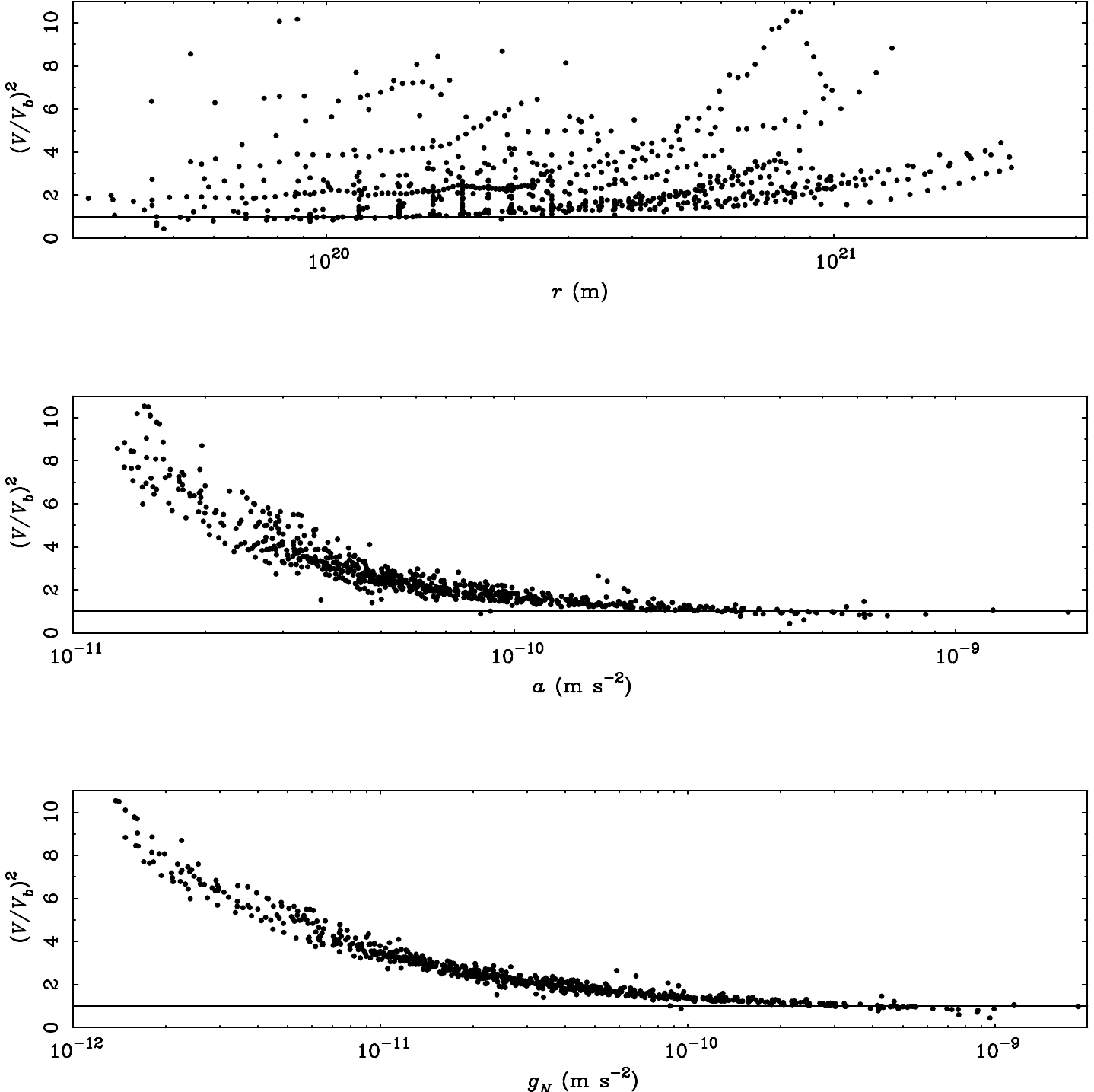}
\caption[Mass discrepancy in disk galaxies]{Mass discrepancy correlates with acceleration (from \citealp{famaey12}). The mass discrepancy is measured as the square of the ratio of $V$ and $V_\mathrm{b}$, where $V$ is the measured rotational velocity and $V_\mathrm{b}$ the velocity calculated from the distribution of the baryonic matter in the Newtonian way. The ratio was measured for a~lot of disk galaxies of many types at various radii. We can see that the mass discrepancy does not correlate with the galactocentric radius, $r$, but it correlates better with the observed acceleration, $a$, and the best with the acceleration calculated by the Newtonian way $g_\mathrm{N}$, in accordance with \equ{algrel}. Note that some additional scatter comes from the uncertainty in the mass-to-light ratio, inclination and the distance of the galaxy from Earth.}
\label{fig:corr}
\end{figure}

A lot of definitions of MOND appeared. They are all motivated by the following correlations found observationally for many types of cosmic objects (see \sect{impl} and \fig{corr}).

\begin{enumerate}
\item Let $\mathbf{a}_\mathrm{N}(\mathbf{r})$ be the gravitational acceleration calculated by the classical Newtonian way from the distribution of the observable matter at the position $\mathbf{r}$. Then observations suggest that a~constant $a_1$ exists, so that the regions of space where
\begin{equation}
	a_\mathrm{N}(\mathbf{r}) \lesssim a_1	
	\label{eq:corr1}
\end{equation}
coincide with the regions where the MMP occurs.
\item Let $\mathbf{a}(\mathbf{r})$ be the observed acceleration of the bodies at the position $\mathbf{r}$.  Let $\mathbf{r}$ be in the region where the MMP appears (i.e., \equ{corr1} is satisfied). Then observations suggest that a~constant $a_2$ exists, so that
\begin{equation}
	a(\mathbf{r}) \approx \sqrt{a_2a_\mathrm{N}(\mathbf{r})}
	\label{eq:corr2}
\end{equation}
and the directions of $\mathbf{a}$ and $\mathbf{a}_\mathrm{N}$ the same within the observational uncertainty.
\item Moreover
\begin{equation}
	a_1 \approx a_2.
	\label{eq:corr3}
\end{equation}
Hence we can set $a_0 = a_1 = a_2$. This constant has the value of $a_0 = (1.24\pm14)\times 10^{-10}$\,m\,s$^{-2}$ \citep{mcgaugh11}. 
\end{enumerate}
On the contrary, we can interpret the correlations \ref{eq:corr1}--\ref{eq:corr3} as an algorithm for predicting the motion of the above-mentioned bodies from the distribution of the observable matter. In this work, we will define MOND as this algorithm. 

Such algorithm could be explained by a~special balance between the distribution of the dark and the observable matter. But we do not have any theoretical explanation why such a~special balance should exist. This behavior is even not reproduced by large cosmological simulations in the DMF. Certain improvement was achieved by postulating the relation between DM halos and stellar wind speeds \citep{illustris}, which is however unphysical \citep{kroupacjp}.  Perhaps the correlations \ref{eq:corr1}--\ref{eq:corr3} will be recovered when the baryonic physics is understood better. In contrast, it is simple to explain these correlations as a~consequence of a~general law of nature -- a~theory of MOND (see \sect{theo}). This would easily explain why the correlations \ref{eq:corr1}--\ref{eq:corr3} are so tight and universal.

\subsubsection{MOND theories}\label{sec:theo}
The modern definition of the MOND theories based on the space-time scaling invariance was presented by \citet{milg09}.  That paper also shows its relation to the older formulations. A~non-relativistic MOND theory is any theory meeting the following tenets:
\begin{enumerate}
	\item It includes a~constant with the dimension of acceleration $a_0$.
	\item Newtonian dynamics is restored in the limit of accelerations much greater than $a_0$ (i.e. the equations reduce to the classical ones when taking the limit $a_0\rightarrow 0$). ``Accelerations'' means all quantities with the dimension of acceleration. This is similar as in General Relativity: Newtonian dynamics is restored if all quantities with the dimension of velocity are small compared to $c$, including, e.g., the square root of the gravitational potential.
	\item For purely gravitating systems, if all accelerations are much lower than $a_0$ (i.e. in the limit $a_0\rightarrow\infty$, $G\rightarrow 0$, keeping the product $Ga_0$ fixed), then the space-time scaling symmetry of the theory emerges: If the equations of the theory are satisfied for a~system of bodies on the trajectories $(\mathbf{r}_i, t)$, then these equations are also satisfied if these bodies move on the  trajectories expanded by a~constant factor in space and time $\lambda(\mathbf{r}_i, t)$, $\lambda>0$. The limit of small accelerations is called the deep-MOND regime.
\end{enumerate}
A lot of predictions follow from these basic tenets alone. Some of them are derived in \sect{impl}; many other can be found in \citet{milgmondlaws}.

As we can see, many MOND theories are allowed to exist \citep{milgcjp}. It is unknown if the right MOND theory should be a~theory of modified gravity, modified inertia, or a~combination of both the options. It comes out that physics is very similar in all MOND theories constructed so far.

Two non-relativistic fully-fledged MOND theories have been developed so far. They are both theories of modified gravity. The first is called AQUAL \citep{bm84}. In this theory, the gravitational potential $\phi_\mathrm{A}$ is given by the equation
\begin{equation}
	\nabla\left[\nabla\phi_\mathrm{A}\,\mu\!\left(\left|\nabla\phi_\mathrm{A}\right|/a_0\right)\right] = 4\pi G\rho
	\label{eq:aqual}
\end{equation}
with the boundary condition $\nabla\phi_\mathrm{N}(\mathbf{r})\rightarrow 0$ for $\mathbf{r}\rightarrow\infty$. The equation of motion reads $-\nabla\phi_\mathrm{A} = \ddot{\mathbf{r}}$. The function $\mu(\cdot)$ is called the interpolating function. It can be chosen arbitrarily, but it has to meet the limit criteria
\begin{itemize}
	\item $\mu(x) \approx 1$ for $x\gg 1$
	\item $\mu(x) \approx x$ for $x\ll 1$. 
\end{itemize} 
Further limitations follow from observations (see \sect{mu}).

Numerical solvers of the nonlinear equation \equ{aqual} use the multigrid methods \citep{brada99, tiret07, raymond} or the multipole expansion in the spherical coordinates \citep{nmody}.

The other non-relativistic MOND theory is QUMOND \citep{qumond}. Its gravitational potential $\phi_\mathrm{Q}$ is the solution of the equation
\begin{equation}
	\Delta\phi_\mathrm{Q} = \nabla\left[\nu\!\left(\left|\nabla\phi_\mathrm{N}\right|/a_0\right)\nabla\phi_\mathrm{N}\right],
	\label{eq:qumond}
\end{equation}
were $\phi_\mathrm{N}$ is the classical Newtonian potential 
\begin{equation}
	\Delta\phi_\mathrm{N} = 4\pi G\rho
\end{equation}
and the function  $\nu(\cdot)$ is related to the interpolating function $\mu(\cdot)$ by the equations
\begin{equation}
	\nu(y) =1/\mu(x), \quad x\mu(x) = y.
\end{equation}
The equations of QUMOND are easy to solve, because we have to only solve the ordinary Poisson equation twice. To calculate the QUMOND potential, numerical codes  use the standard adaptive mesh refinement multigrid methods  \citep{angus12, raymond, por}.

Modified inertia MOND theories are allowed to exist \citep{milg94, milg06, milg11, milgcjp}, but no one has been constructed so far. In the modified inertia theories, the acceleration of a~body depends on its whole trajectory from the beginning of the Universe, should some standard theoretical requirements hold true \citep{milg94}. For this reason, we cannot define the gravitational potential in them. \citet{milg12b} suggested an observable parameter that could be used to discern between the MOND modified gravity and modified inertia theories. We could also discriminate between various modified inertia theories and modified gravity theories by observing bodies on different shapes of trajectories (\citealp{bil14}, see also \sect{mondsh}).

A number of MOND relativistic theories exist. See \citet{famaey12} for a~thorough review. They include TeVeS \citep{teves}, BIMOND \citep{bimond} or a~subset of Einstein-Aether theories \citep{zlosnik}, to name a~few.

For the practical purposes, we often resort to the original (also called pristine) formulation of MOND \citep{milg83a, milg83b, milg83c}. It states that in the gravitational field, bodies move so that the equation (called the algebraic relation)
\begin{equation}
	\mathbf{a}\,\mu\!\left(\frac{a}{a_0}\right) = \mathbf{a}_\mathrm{N}
	\label{eq:algrel}
\end{equation}
holds true\footnote{In the pristine formulation, the equation of motion reads $\mathbf{a}\mu(a/a_0) = \mathbf{F}$, which was interpreted as a~modification of the left-hand-side of Newton's equation of motion $\mathbf{a} = \mathbf{F}$ by \citet{milg83a}, i.e. the reaction of a~body on the exerted force is modified. For this reason, the word MOND originally stood for MOdified Newtonian Dynamics. Since MOND can also be interpreted as a~modification of gravity or as a~balance between the observable and dark matter, the word MOND is no more considered an abbreviation.}. Note the similarity\footnote{The difference is that this equation states a~precise equality (not only a~correlation) and, in the original formulation, the algebraic relation is, strictly speaking, universal for all bodies (the algorithm is valid only for certain types of objects).} to the observational MOND algorithm represented by Eqs.~\ref{eq:corr1}--\ref{eq:corr3}.

We should emphasize that the only observation which inspired the original formulation (and subsequently all the MOND theories) was the flattens of rotation curves at large radii. All the other consequences of the MOND theories listed in \sect{impl} are predictions, which get  confirmed in most cases.

The original formulation is not a~theory \citep{milg83a}. For example, it does not conserve momentum (consider the gravitational attraction of two bodies with different masses) or we encounter problems when calculating the motion of composite bodies (a star orbiting a~galaxy consists of particles moving with accelerations higher than $a_0$, but the star orbits the galaxy with an acceleration lower than $a_0$).

However, both in AQUAL and QUMOND theories the algebraic relation is precise for a~test particle moving in a~gravitational field with the spherical, cylindrical (called axial in the original papers) or translational (infinite planes stacked on top of each other) symmetry, as we can see by applying Gauss's theorem \citep{bm84,qumond}. Furthermore, it is precise for several special mass distributions \citep{brada95}, most notably for the Kuzmin disk. The algebraic relation is also precise for bodies on circular orbits in the modified inertia theories \citep{milg94}. The gravitational fields resulting from AQUAL and the algebraic relation were compared by \citet{brada95} and \citet{ciotti06} on a~few examples.

\subsubsection{The interpolating function}\label{sec:mu}
The two most widely used interpolating functions are the ``standard'' interpolating function
\begin{equation}
	\mu(x) = \frac{x}{\sqrt{1+x^2}}
	\label{eq:mustand}
\end{equation}
and the ``simple'' interpolating function
\begin{equation}
	\mu(x) = \frac{x}{1+x}.
	\label{eq:musimp}
\end{equation}
Many more interpolating functions appeared in literature. It is believed that the correct form of the interpolating function will follow from a~deeper theory or it will come out that no universal interpolating function exists. The attempts to derive the interpolating function up to now include \citet{milg99, klinkhamer11}, or \citet{trippe13}. \citet{famaey07} claim that the simple interpolating function works best for the fitting of rotation curves. We know that the simple and standard interpolating function do not work for high accelerations because they approach the Newtonian regime too slowly \citep{milg83a}. Further observational constraints were discussed in \citet{famaey07} and \citet{famaey12}.

\begin{figure}[ht!]
	\centering\includegraphics[width=0.6\textwidth]{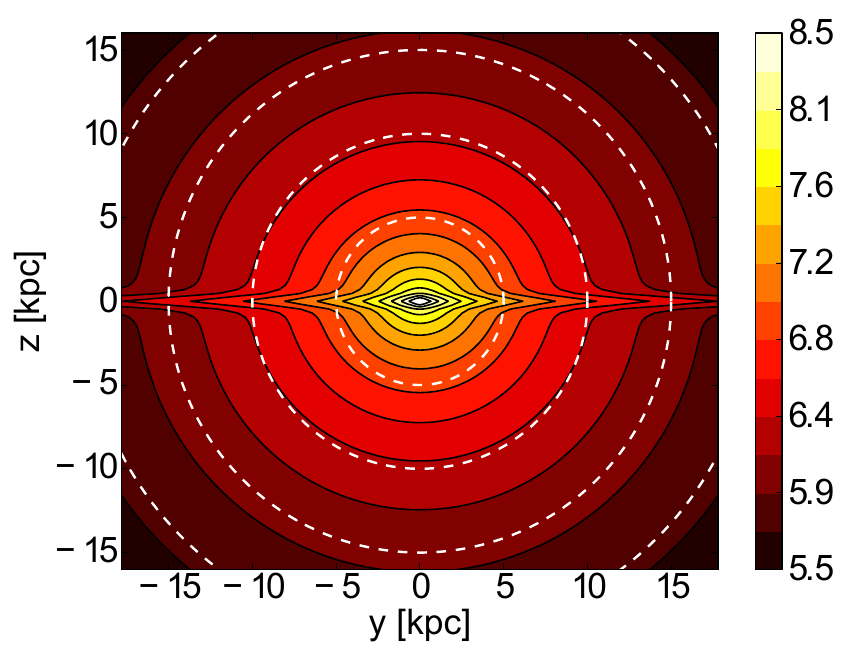}
\caption[Phantom dark matter in a disk galaxy]{Density of the phantom dark matter in a~disk galaxy (from \citealp{por}). }
\label{fig:phantom}
\end{figure}

\subsubsection{Phantom dark matter}
The works on MOND often use the term ``phantom dark matter'' (PDM). It is a~mathematical construction advantageous for comparing the DMF with MOND. It is the DM we need to add to the observable matter in the Newtonian dynamics, so that the resulting gravitational field is the same as the gravitational field calculated by MOND. If $\phi_\mathrm{M}$ is the MOND potential, then the density of the PDM can be calculated as
\begin{equation}
	\rho_\mathrm{PDM} = \frac{1}{4\pi G}\Delta\phi_\mathrm{M} - \rho,
	\label{eq:pdm}
\end{equation}
where $\rho$ is the density of the observable matter. As an example, the distribution of the PDM in a~disk galaxy is shown in \fig{phantom}.

\subsubsection{Numerical coincidences}
The MOND acceleration constant $a_0$ is a~subject of various numerical cosmological coincidences. It was noticed already by \citep{milg83a} that $2\pi a_0 \approx cH_0 \approx c^2\sqrt{\Lambda/3}$ where $H_0$ is the Hubble constant and $\Lambda$ the cosmological constant. The combination $ \frac{c^4}{Ga_0}\approx \frac{2\pi c^3}{GH_0}\approx  \frac{2\pi c^2}{G(\Lambda/3)^{1/2}} \approx 10^{54}$\,kg is the total mass of the observable part of the Universe. The length $\frac{c^2}{a_0}\approx 7.5\times10^{30}$\,m is of the order of the current Hubble radius. It is unknown if these coincidences have a~deeper reason. In the bimetric formulation of MOND (one of the MOND relativistic theories), the coincidence between $a_0$ and $\Lambda$ is a~direct implication of the theory \citep{bimond}.

\subsection{ MOND predictions and theoretical results}\label{sec:impl}
Here we list some of the theoretical results about the MOND theories and their observable predictions. Many of these implications follow from the basic tenets so that they are independent of the particular MOND theory. 

\begin{itemize}

\begin{figure}[ht]
  \begin{minipage}[c]{0.45\textwidth}
    \includegraphics[width=\textwidth]{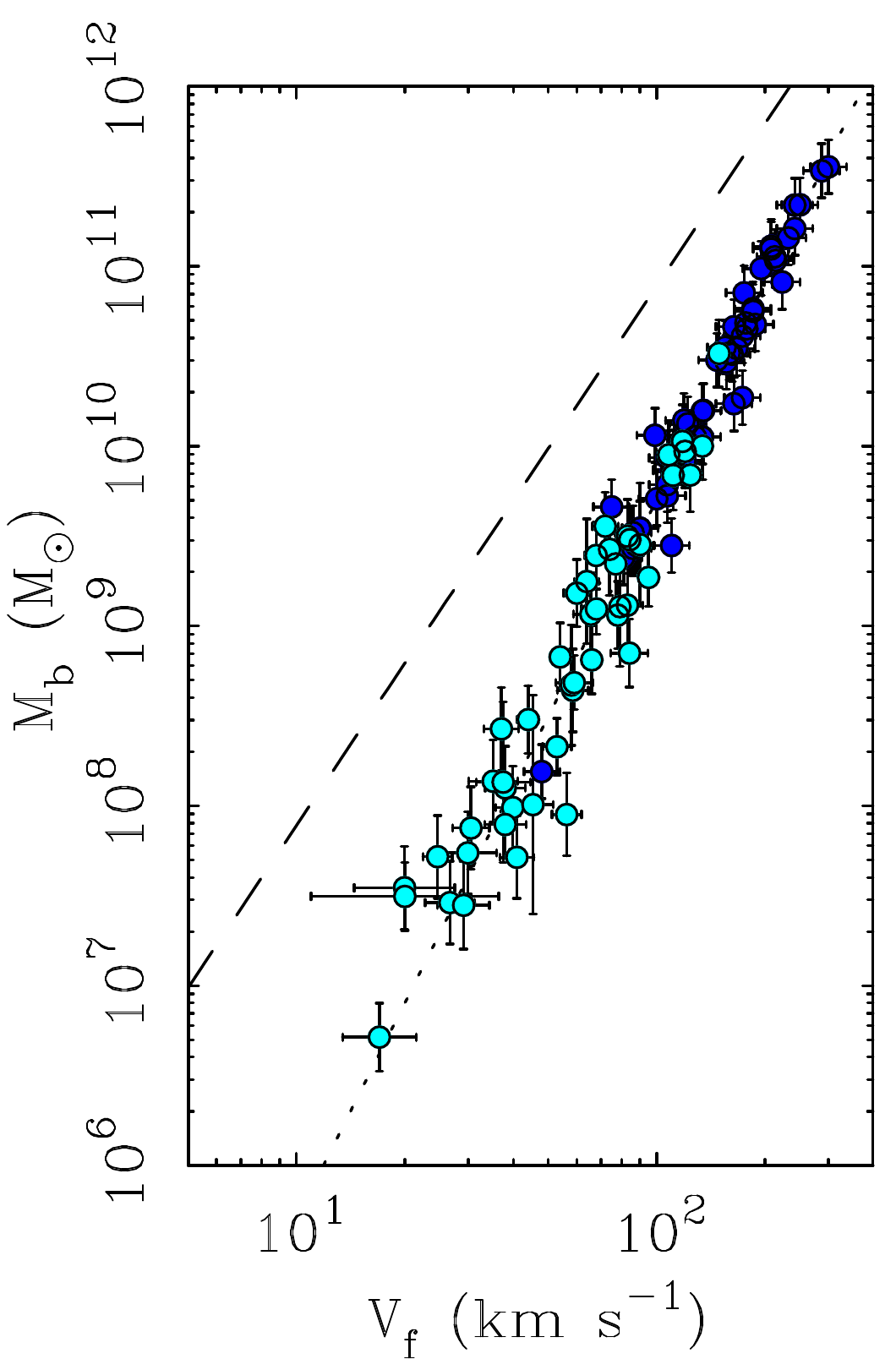}
  \end{minipage}\hfill
  \begin{minipage}[c]{0.52\textwidth}
    \caption[Baryonic Tully-Fisher relation]{Baryonic Tully-Fisher relation (from \citealp{famaey12}). Horizontal axis -- Terminal rotational velocity. Vertical axis -- Total baryonic mass of the galaxy. Light blue points -- Galaxies containing more mass in the gas than in the stars. Dark blue points -- Galaxies dominated by stars. Dotted line -- Baryonic Tully-Fisher relation predicted by MOND (see \equ{btfr}). Dashed line -- A~simplified prediction by the DMF supposing a~universal ratio of the dark and baryonic mass of every galaxy.
        } \label{fig:btfr}
  \end{minipage}
\end{figure}

	\item Rotation curves of isolated galaxies in MOND get flat at very large radii, i.e. the rotation speed does not depend on radius (theoretical derivation in the pristine formulation: \citealp{milg83a}, derivation in the modern formulation: \citealp{milgmondlaws}, the original observation: \citealp{rubin80}). The flatness of galactic rotational curves at large radii is not a~prediction. MOND was constructed to account for this observation \citep{milg83a}.  However, the orbital velocities of wide binary stars also appear to be independent of the separation of the components \citep{hernandez12}.
	
	\item The terminal velocity $V_\infty$ is given by the total baryonic mass of the galaxy $M$ as 
	\begin{equation}
		V_\infty = \sqrt[4]{GMa_0}.
	\label{eq:btfr}
\end{equation}
 This is called the baryonic Tully-Fisher relation (derivation in \citealp{milg83a} and \citealp{milgmondlaws}; observational confirmation, e.g., in \citealp{mcgaugh11}). The basic tenets dictate only the proportionality between $M$ and $V_\infty$, but the normalization $a_0$ must be determined observationally so that \equ{btfr} holds true.

To demonstrate how the basic tenets of MOND implicate the results independent of the MOND theory, we will derive \equ{btfr}. Consider a~test particle orbiting a~galaxy of the total mass $M$ on a~circular trajectory with a~very large radius $r$ (i.e. the centripetal acceleration is small at $r$ compared to $a_0$, and the vast majority of the galaxy mass is accumulated under $r$). The velocity at large radii, $V_\infty$, can depend only on the quantities $M$, $r$, $G$ and $a_0$. From the dimensional grounds, $V_\infty$ has to have the form of $V_\infty = \sqrt{a_0r}f\!\left(\frac{GM}{a_0r}\right)$, where $f$ is a~function which cannot be recovered by the dimensional analysis. Now, preparing to employ the scaling invariance, we substitute $\lambda \mathbf{r} \rightarrow \mathbf{r}$ and $\lambda t \rightarrow t$. The velocity does not change because $\frac{\d \lambda \mathbf{r}}{\d \lambda t} = \frac{\d \mathbf{r}}{\d t}$, so we have $V_\infty = \sqrt{a_0\lambda r}f\!\left(\frac{GM}{a_0\lambda r}\right)$ for the expanded orbit. To meet the requirement of the scaling invariance (to cancel out $\lambda$), the function $f$ has to have the form of $f(x) = c\sqrt{x}$, where $c$ is an arbitrary constant. In total, we have $V_\infty = c\sqrt[4]{GMa_0}$. The constant $a_0$ is set so that $c=1$ for the circular trajectories. Once $a_0$ is fixed, the value of $c$ can differ with the shape of the orbit.

  \item Isolated finite objects produce gravitational acceleration $a(r) = \sqrt{GMa_0}/r$ and the potential $\phi(r) = \sqrt{GMa_0}\ln(r)+\phi_0$ at large radii (assuming a~modified gravity theory). This follows from the previous finding. 
	
	\item Mass discrepancy appears in a~galaxy at the radius where the observed acceleration drops below $a_0$. This radius is called the transitional radius \citep{milg83a, milgmondlaws}.  The transitional radius for a~point mass is $r_\mathrm{M} = \sqrt{GM/a_0}$, which is a~good approximation for most galaxies.

	\item Rotation curves meet 
	\begin{equation}
	 \frac{v_\mathrm{c}^2(r)}{r}\,\mu\!\left[\frac{v_\mathrm{c}^2(r)}{a_0r}\right] = \frac{GM(r)}{r^2},
	\end{equation}
	where $M(r)$ is the cumulated baryonic mass under the radius $r$ and $v_\mathrm{c}(r)$ the circular velocity at the same radius. This equation is precise in the modified inertia theories \citep{milg94} and only approximate in the modified gravity theories \citep{brada95} but it gets precise for large radii to reproduce the universal \equ{btfr}. The rotation curves calculated from the distribution of the baryonic matter by this way reproduce even small details on the observed rotational curves for many galaxies (see examples in \citealp{famaey12} and \fig{rc}).

	\item The rotation curves of tidal dwarf galaxies are difficult to be explained in the DMF \citep{bournaud10}. These galaxies are self-bound objects formed in tidal arms of interacting disk galaxies. In the DMF simulations, only little tidal dwarfs form and they are devoid of DM. But to explain the rotation curves of the observed tidal dwarfs, DM is needed supposing the DMF, see \fig{tidal_rotcurve}. These galaxies do not pose a~problem for MOND \citep{gentile07}. They form easily in simulations and their observed rotation curves are compatible with the theory, see \fig{tidalDMMD}.

\begin{figure}[t]
	\centering\includegraphics[width=0.9\textwidth]{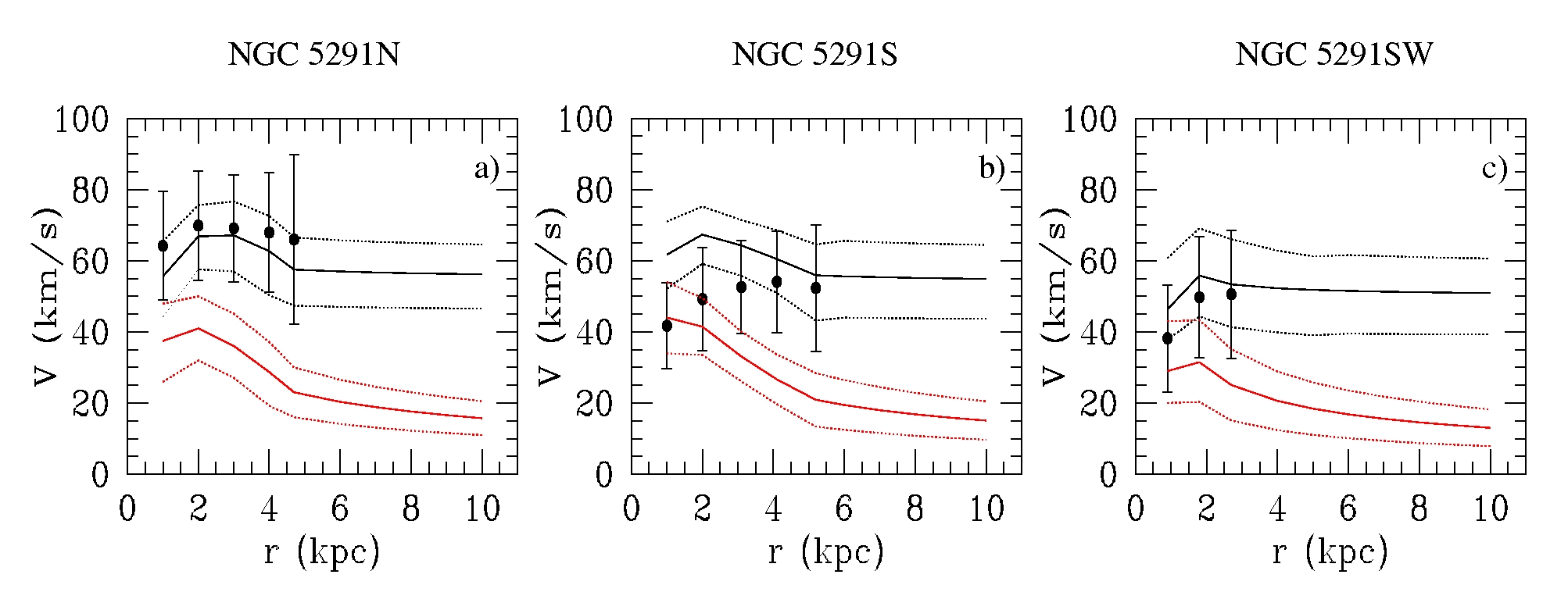}
\caption[Rotation curves of tidal dwarf galaxies]{Rotation curves of tidal dwarf galaxies (from \citealp{famaey12}). Black limits --  Rotation curves predicted by MOND. Red limits -- Rotation curves predicted by the DMF (no DM is expected in tidal dwarfs for theoretical reasons).}
\label{fig:tidal_rotcurve}
\end{figure}

\begin{figure}
	\centering\includegraphics[width=0.9\textwidth]{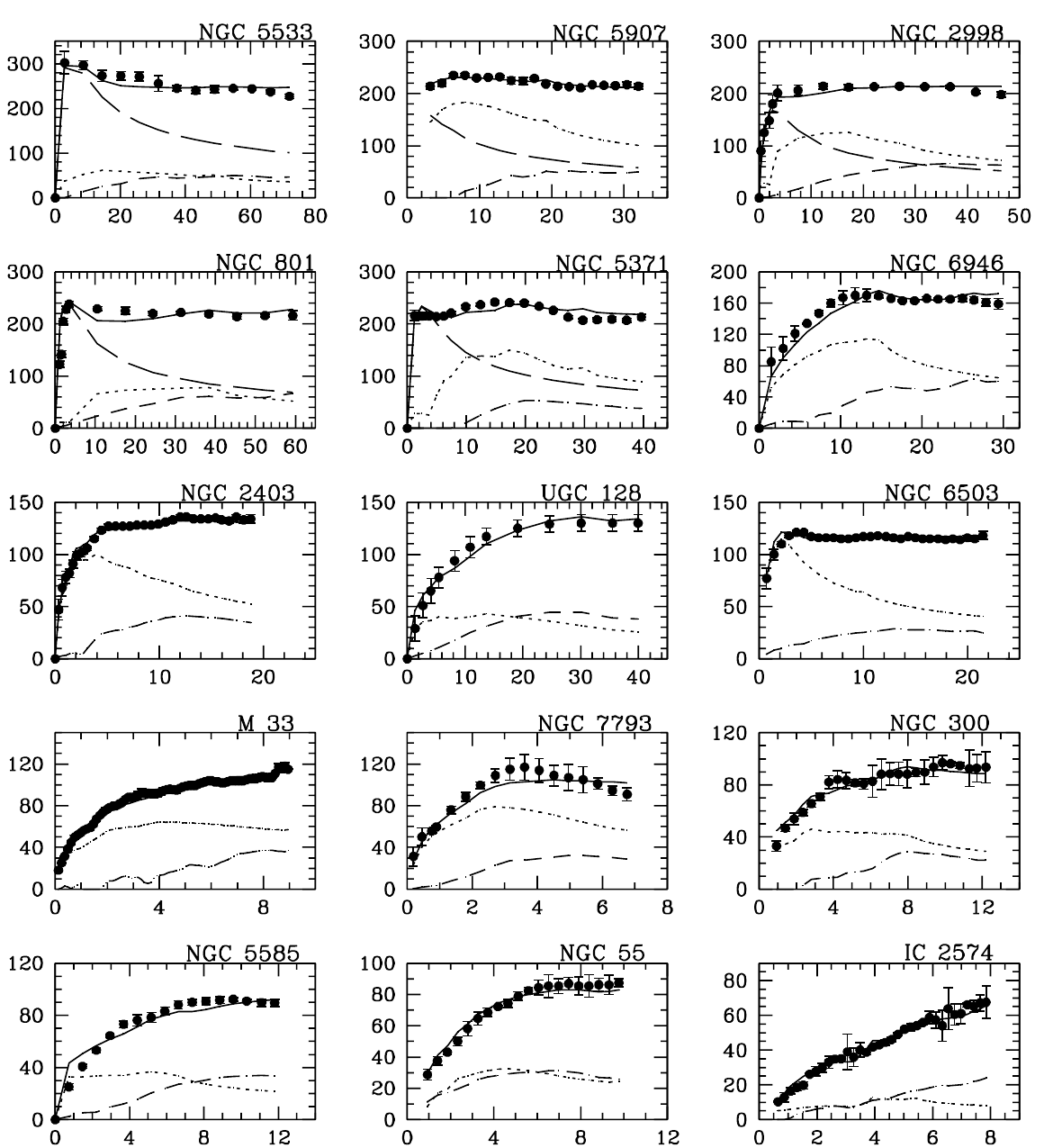}
\caption[MOND fits of rotation curves]{MOND fits to rotation curves of several disk galaxies (from \citealp{sanders02}). The radius is given in kiloparsecs and the rotational velocity in kilometers per second. The solid curve is the MOND fit. Its only free parameter was the mass-to-light. Dotted line -- rotation curves calculated in the Newtonian way from the stellar component. Dashed line -- The same for the gaseous component. }
\label{fig:rc}
\end{figure}

\begin{figure}
	\centering\includegraphics[width=0.5\textwidth]{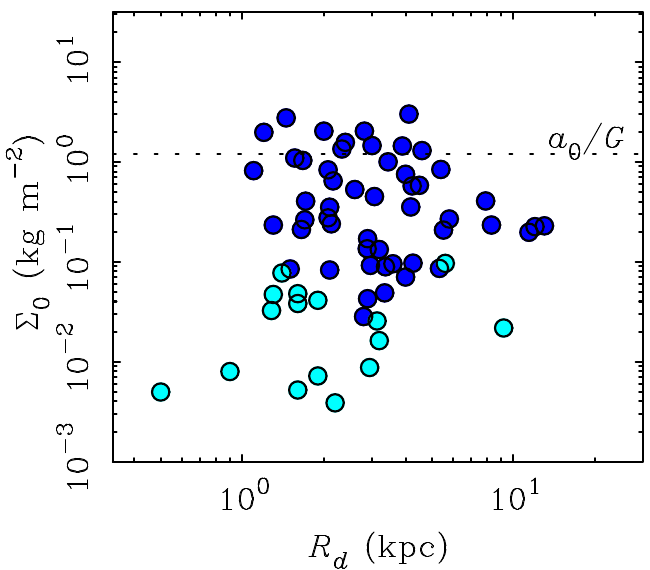}
\caption[Freeman limit]{ Freeman limit (from \citealp{famaey12}).  Each point represents a~disk galaxy. The surface density of the disk component was fitted by the exponential profile $\Sigma(r) = \Sigma_0\exp\left(-r/R_\mathrm{d}\right)$. The plot shows that galaxies are rare above a~certain value of surface density -- the Freeman limit. The dashed line is the value the Freeman limit predicted by MOND ($\Sigma_\mathrm{crit} = a_0/G$). }
\label{fig:freeman}
\end{figure}

	\item Galaxy discs  have increased stability in the deep-MOND regime than in Newtonian dynamics without the DM \citep{brada99}. Disk galaxies are unstable in the strong acceleration regime \citep{ostriker73} without the DM halos. This explains the observational Freeman limit stating that disk galaxies having surface brightness higher than a~certain value are rare (e.g., \citealp{mcgaugh96}), see \fig{freeman}. According to MOND, this limit is $\Sigma_\mathrm{crit} = a_0/G$. Disk stability with respect to the Toomre Q~parameter of MOND was tested observationally by \citet{jimenez14}. 

  \item The distribution of PDM halos in disk galaxies have a~spherical and disk component, i.e. if a~MOND theory applies but we use the Newtonian dynamics, we always derive an increased density of DM in the plane of the disk (\citealp{milg01, milgmondlaws}, see also \fig{phantom}).

	\item  The characteristic three-dimensional velocity dispersion of the pressure supported systems, like globular clusters, elliptical galaxies and galaxy clusters, is given by 
	\begin{equation}
		\sigma^2 = Q (GMa_0)^2,
	\label{eq:sigma}
\end{equation}
 where the constant $Q$ depends on the theory, but it is of the order of unity \citep{milg84, milgmondlaws}. In all modified gravity MOND theories, this constant is $Q = 2/3$ for isolated isothermal spheres in the deep-MOND regime \citep{milg14}. For the real objects, \equ{sigma} has only an approximate validity because they are not isothermal spheres, they contain the high-acceleration region in the center, etc. \citep{famaey12}. This relation was first recognized observationally as the Faber-Jackson relation \citep{faber76}, see \fig{fabjac}. It holds true for the pressure supported systems from globular clusters to galaxy clusters \citep{sanders10, famaey12}. Velocity dispersion profiles approach asymptotically the value given by \equ{sigma}, as was tested, e.g.,  by \citet{scarpa11} or \citet{hernandez12}.

	\begin{figure}[t]
	\centering\includegraphics[width=0.5\textwidth]{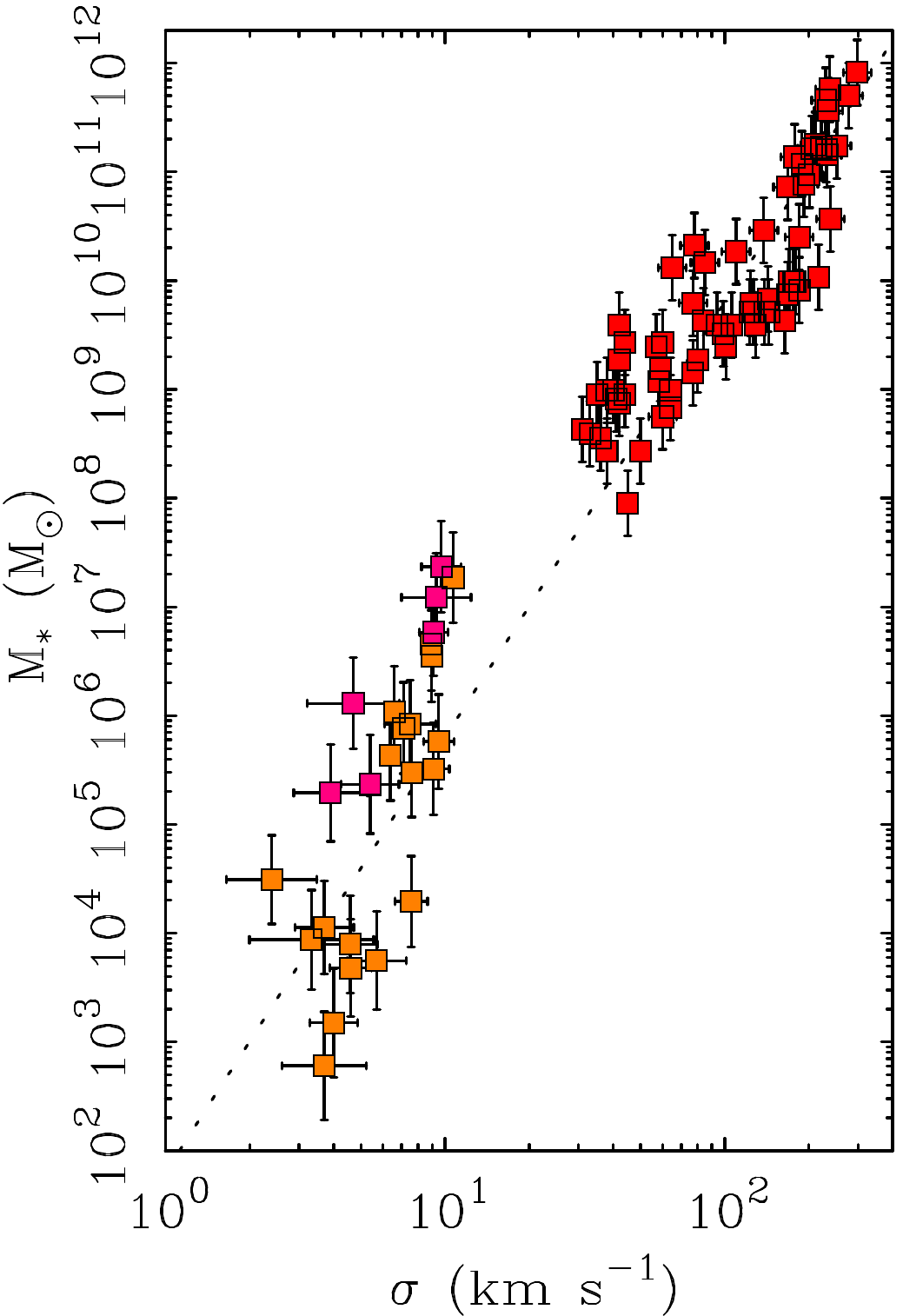}
\caption[Faber-Jackson relation]{Faber-Jackson relation for giant and dwarf elliptical galaxies (from \citealp{famaey12}). The stellar mass of the galaxy is $M_\star$ and the line-of-sight velocity dispersion is $\sigma$. The dashed line is the MOND prediction given by \equ{sigma} (no fitting here). }
\label{fig:fabjac}
\end{figure}

	\item The density of galactic stellar halos drops as $\rho\propto r^{-3}$ \citep{hernandez13}.

	\item The the acceleration caused by the PDM calculated by the Newtonian way can never much exceed $a_0$ \citep{brada99b, milgmondlaws}. 

	\item Rings of dark matter deduced via gravitational lensing studies in some galaxy clusters \citep{jee07} are explained naturally by MOND. The typical surface density of such PDM ring deduced using the conventional dynamics is predicted to be around $\Sigma_0 = a_0/G$ \citep{milg08}.

	\item The density of the PDM can be negative for certain distributions of the baryonic matter \citep{milg86}. Such regions occur, e.g., between an isolated pair of galaxies. 

	\item Gravitational lensing is in accordance with observations for some relativistic extensions of MOND (see \citealp{famaey12} for a~review). To calculate the trajectory of a~photon in all the known relativistic versions of MOND, we can use the equations of General Relativity if we treat the PDM as DM \citep{milg13}. The deflection angle gets asymptotically independent of radius for an isolated point mass \citep{mortlock01}. The asymptotic deflection angle is $\Delta\alpha = -2\pi \sqrt{GMa_0}/c^2$.

	\item All acceptable MOND theories are non-linear \citep{milgmondlaws}, i.e. if we have a~particle moving in the field produced by the density $\rho_1+\rho_2$, then this motion cannot be obtained by the superposition of the motions calculated from $\rho_1$ and $\rho_2$ separately. The basic tenets of MOND allow for a~linear theory, but this theory is not compatible with observations.
\end{itemize}

The dynamics of a~system is affected by the external acceleration field in MOND. It is a~consequence of the non-linearity of MOND. The external field effect (EFE) is different from tidal forces. For example, if a~globular cluster moves in a~homogeneous external gravitational field, the internal motions of the stars in the cluster are affected by the external field.  In the Newtonian dynamics, the cluster would fall as a~whole in such field and the internal motions would be the same as if the external field was zero.

\begin{itemize}
	\item If the internal gravitation field of a~system $a_\mathrm{int}$ is stronger than $a_0$, the system behaves in the Newtonian way irrespective of the magnitude of the external field $a_\mathrm{ext}$ \citep{milg83a, milgmondlaws}. 

	\item If $a_\mathrm{ext}<a_\mathrm{int}<a_0$, the system behaves in the MOND way \citep{milg83a, milgmondlaws}.

	\item If $a_\mathrm{int}<a_\mathrm{ext}< a_0$, then the internal dynamics is quasi-Newtonian, i.e. Newtonian with the value of the gravitational constant increased by the factor of $a_0/a_\mathrm{ext}$ \citep{milg83a, milgmondlaws}. This effect was detected for the satellites of M\,31 \citep{anddwarf, anddwarfii} and also for the Milky Way satellites with some tension, though \citep{lughausen14}.

	\item If $a_\mathrm{ext}>a_0$, then the system behaves in the Newtonian way \citep{milg83a, milgmondlaws}.

\begin{figure}[t]
	\centering\includegraphics[width=0.5\textwidth]{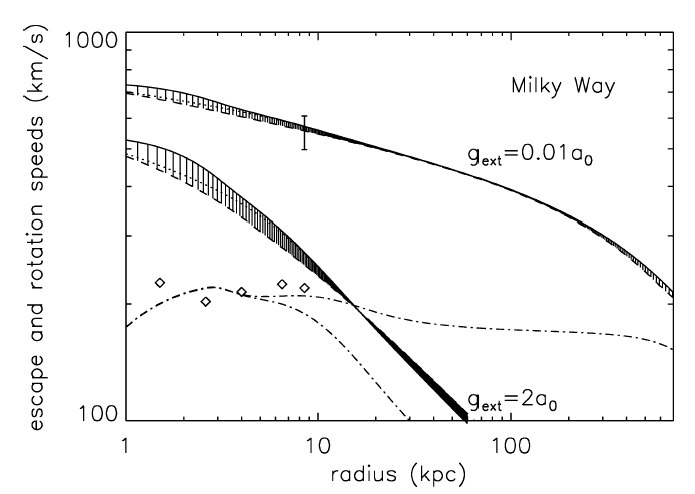}
\caption[External field effect in Milky Way]{External field effect (from \citealp{wu07}). Shaded areas -- Escape velocity from Milky Way calculated for MOND as a~function of the galactocentric radius for the external field intensity of $0.01a_0$ (upper area) and $2a_0$ (lower area). The data point shows the the escape velocity in the neighborhood of Sun by the RAVE experiment. Dash-dot lines --  Rotation curve of Milky Way in MOND calculated for the external field of $0.01a_0$ (upper curve) and $2a_0$ (lower curve). }
\label{fig:escape}
\end{figure}

	\item  If a~system gets from a~weak external field into a~strong internal field, like a~satellite falling on a~big galaxy, and the system was originally in the deep-MOND regime,  then the system loses stability and disintegrates \citep{satelliteefe}. In a~Newtonian analogy, it is like if the DM halo of the satellite disappeared. On the contrary, if the satellite or a~globular cluster moves outward in the galaxy, it contracts when the external field drops below $a_0$ \citep{wu13}.

	\item The internal acceleration field of a~spherically symmetric body falling in a~homogeneous external field is not spherically symmetric \citep{wu13}. 

	\item Galactic discs can warp by the EFE from their satellites \citep{brada00}. 

	\item Real galaxies move in the external field of the neighboring galaxies and cosmogical structures.  Far enough from any real galaxy, this large-scale field prevails over the field from the galaxy, so that the acceleration from the galaxy switches to the quasi-Newtonian regime and declines like $a\propto r^{-2}$. This means that rotation curves are not flat to infinity even in MOND, but have a~Keplerian decline at large radii caused by the EFE (see more details in \citealp{wu07} and \fig{escape}).

	\item The escape velocity from a~galaxy is defined by the magnitude of the external field. Isolated object have logarithmic potential in MOND, so their escape velocity is infinite, but such objects do not exist. The external field exerted on Milky Way is around $0.01a_0$ and is caused mainly by the Andromeda galaxy and Virgo and Coma Clusters \citep{famaey07b}. The observed escape velocity from Milky Way is in accordance with the MOND expectation \citep{famaey07b, wu07}, see \fig{escape}. If a~disk galaxy arrives to a~sufficiently strong external field, the escape velocity can decrease so much that the stars can unbound from the outskirts of the galaxy and fly away. 

	\item The EFE can be negligible in some MOND modified inertia theories for some trajectories \citep{milg11}.
\end{itemize}

\begin{figure}[t]
	\centering\includegraphics[width=0.9\textwidth]{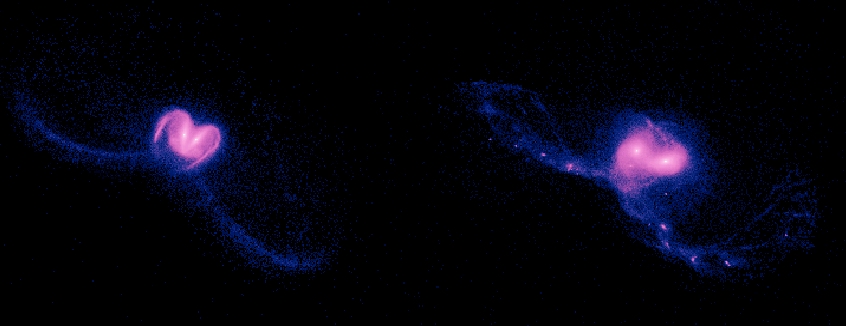}
\caption[Interacting disk galaxies in MOND and the dark matter framework]{Simulation of interacting disk galaxies in the DMF (left) and MOND (right) (from \citealp{combtir10}). Note the formation of tidal dwarf galaxies in the MOND simulation and their lack in the DMF simulation. }
\label{fig:tidalDMMD}
\end{figure}

\subsection{Galactic simulations in MOND}\label{sec:mondsim}
The following results were obtained by simulations in the AQUAL and QUMOND theories of MOND. It is unclear if they would come out similarly in other MOND theories, especially in the modified inertia theories. The simulations comparing MOND with the DMF start from the so-called equivalent systems. This means that we first define the distribution of the baryonic matter and calculate the MOND gravitational field. We obtain the distribution of the PDM, which is fitted by a~standard DM halo. This halo is then used in the Newtonian simulation.

\begin{itemize}
	\item The disk and interacting galaxies simulated in MOND show the observed morphologies (disc galaxies: \citealp{tiret07, tiret08b, por}, interacting galaxies: \citealp{tiret08}), see \fig{galmorf} and \fig{tidaltiret}.

	\item  Dynamical friction is usually weaker in MOND than in the DMF \citep{tiret08, nipoti08}. It is stronger only if the mass of the perturbation constitutes less than 5\% of the total mass of the system \citep{nipoti08}.

\begin{figure}
	\centering\includegraphics[width=0.7\textwidth]{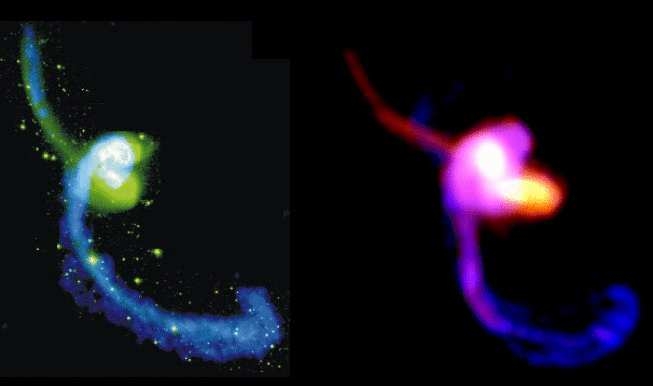}
\caption[Simulation of Antennae galaxies in MOND]{Observation (left) and simulation (right) of the Antennae interacting galaxies in MOND (from \citealp{tiret08}).}
\label{fig:tidaltiret}
\end{figure}

\begin{figure}
	\centering\includegraphics[width=1\textwidth]{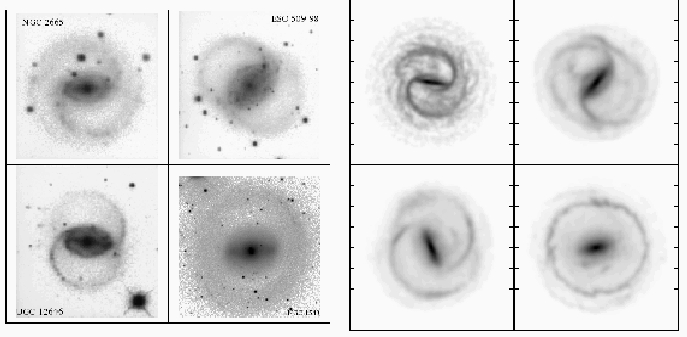}
\caption[Comparison of observed and simulated galaxies in MOND]{ Galaxies simulated in the AQUAL theory of MOND (see \sect{theo}) have the observed morphologies (from \citealp{combtir10}). Left -- Observed galaxies. Right -- Simulated galaxies.}
\label{fig:galmorf}
\end{figure}

	\item The formation of galactic bars is faster in MOND than in the DMF. In the simulations of \citet{tiret07}, the bar strength decreases after some time in MOND, while the bar only grows in the DMF. However, the evolution of the bar depends on the size of the bulge \citep{por}. The preliminary MOND simulations of \citet{tiret07} reproduce the statistical distribution of galaxy morphologies better than the DMF. They contained only stars, though.

	\item In MOND, the pattern rotation speed of a~bar does not decrease with time unlike in the DMF \citep{tiret07}. In the DMF, DM absorps the angular momentum of the bar. Observations show that the average bar rotation speed does not evolve with redshift \citep{perez12}.

	\item The weak dynamical friction in MOND blocks one of the mechanism of bulge formation in disk galaxies \citep{combes14}. In the early stages of the Universe, disk galaxies contain massive gas clouds. In the DMF simulations, these clouds lose energy by dynamical friction and settle to the galaxy center. Here they turn into the stars forming the bulge. In MOND, these clouds lose the orbital energy too slowly, so that they turn into stars before they reach the  galaxy center. The DMF simulations have problems to produce bulgeless galaxies, but up to 80\% of local galaxies above a~stellar mass of $10^9$\,M$_\odot$ contains no bulge or a~pseudobulge \citep{fisher11}.

	\item Galaxy mergers are less frequent and take longer time in MOND because dynamical friction is weaker \citep{tiret08}.

	\item A~fully-self consistent picture of the history of the Local group of galaxies exists in MOND based on the finding that the dynamical friction is lower in MOND than in the DMF \citep{zhao13}. It explains the alignment of the satellites of Milky Way and M\,31 into rotating disk-like structures by their close approach 7-11\,Gyr ago. This is consistent with the orbital times of the satellites, and with the ages of the stellar populations in the satellites and the stellar halos of the big galaxies. The formation of the discs of satellites is problematic for the DMF (see \citealp{pawlowski15} for a~review). 

	\item Rotation curves of galactic polar rings in simulations show the observed qualitative properties \citep{lughausen13}.
\end{itemize}

\subsection{The shortcomings of MOND}\label{sec:mondprob}
\begin{itemize}
	\item 
The most serious shortcoming of MOND is probably \textbf{the remaining missing mass problem} (see \citealp{angus08} for a~review). There is a~missing mass problem even in MOND in the centers of galactic clusters, cD galaxies and some galaxy groups. The problem is usually encountered in X-ray bright systems. The required amount of the missing mass is comparable to the amount of the observable mass. This means either that MOND is incorrect or some matter indeed still escapes observations. 
	
	Two solutions were suggested. The first were the sterile neutrinos. They initially appeared as a~promising idea. They were able to solve the mass discrepancy in galaxy clusters and to reproduce the  power spectrum of the cosmic microwave background \citep{angus08, angus09}. However, MOND cosmological simulations with sterile neutrinos showed that too many too massive galaxy clusters are formed \citep{angus13b}.
	
	The other solution suggests the existence of unseen baryonic matter in the form of small dense non-luminous objects (cluster baryonic dark matter, CBDM, \citealp{milgcbdm}). Their existence could moreover explain another unrelated mystery -- the cooling flow problem. This seems to be the most acceptable solution of the missing mass problem of MOND today.

\begin{figure}
	\centering\includegraphics[width=0.7\textwidth]{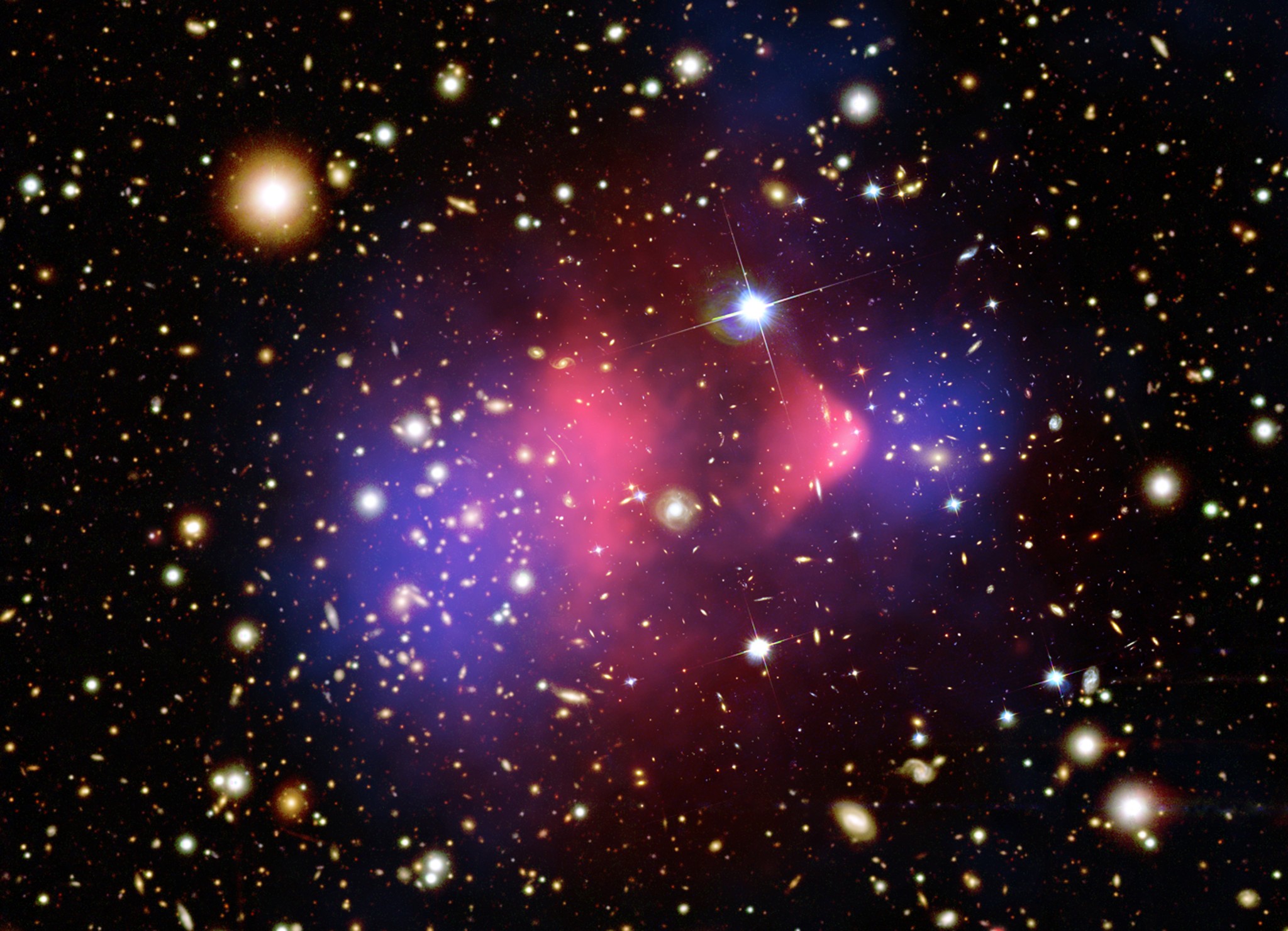}
\caption[Bullet Cluster]{Image of Bullet Cluster (image credit Douglas Clowe and Maxim Markevitch). Red clouds -- Surface density of the hot gas. Blue clouds -- Surface density of DM supposing the DMF.  }
\label{fig:bullet}
\end{figure}

\item
  Several \textbf{relativistic extensions of MOND} have been constructed, but none of them meets all the observational constraints (power spectrum of the cosmic microwave background, gravitational lensing, integrated Sachs-Wolfe effect, a~sufficiently fast approach to General Relativity for high accelerations). See \citet{famaey13} for a~review. 
	
	Since a~reliable relativistic version of MOND is missing, it is difficult to make MOND predictions where a~relativistic theory is required. Those are the areas related to cosmology, where $\Lambda$CDM excels. MOND provides no information here.

\item
Many authors found MOND inconsistent with the velocity-dispersion profiles of \textbf{globular clusters} while others arrived to the opposite result, see \citet{frank14} or \citet{hernandez12} for a~review.

\item
Similarly, the \textbf{rotation curves} of some disk galaxies were found inconsistent with MOND (e.g., \citealp{bottema15, iocco15}). Some of these curves were reanalyzed later with the opposite result (e.g., \citealp{engelke15}). But many more rotation curves exist that are perfectly consistent with MOND \citep{famaey12}. Given that a~lot of factors can affect the appearance of the rotational curve (galaxy interactions, bars, non-circular orbits, a~variable mass-to-light ratio, observational errors, \ldots), it is not surprising that some rotation curves would deviate from the MOND expectations if MOND was a~law of physics. A~more serious problem seems to be the inconsistency of rotation curves with heights of galactic discs \citep{angus15}.

\item
 \textbf{Bullet Cluster} is an object traditionally claimed to be a~direct evidence of exotic DM \citep{clowe06}. It is  pair of galactic clusters seen shortly after their collision. The collisional intergalactic gas decoupled from the collisionless component, i.e. the galaxies, by ram pressure (see \fig{bullet}). The gas is offset from the two gasless galaxy clusters toward the center of mass.  Most mass of the observable matter is in the gas. The analysis of the weak and strong gravitational lensing revealed that the highest concentration of the gravitating matter is located among the galaxies. This is exactly what we expect from DM. If only the observable matter was present in the galaxy, the highest concentration of matter deduced from gravitational lensing is naively expected to be in the gas cloud in the modified gravity theories. But, for example, the MOG modified gravity theory does not need DM in Bullet Cluster at all \citep{brownstein07}. In MOND, some DM is needed but this does not pose an additional problem for MOND as we know that DM is needed in all galaxy clusters in MOND. Moreover, other colliding galaxy clusters exist where the gravitating mass has a~similar distribution as the baryonic mass (e.g., the Train Wreck cluster, \citealp{trainwreck}). Bullet cluster is a~problematic object even for the DMF model because the relative velocity of the colliding clusters is too high \citep{lee10}.
\end{itemize}

\section{Shell galaxies}\label{sec:sh}
\begin{figure}[t]
 \centering\includegraphics[width=0.6\textwidth]{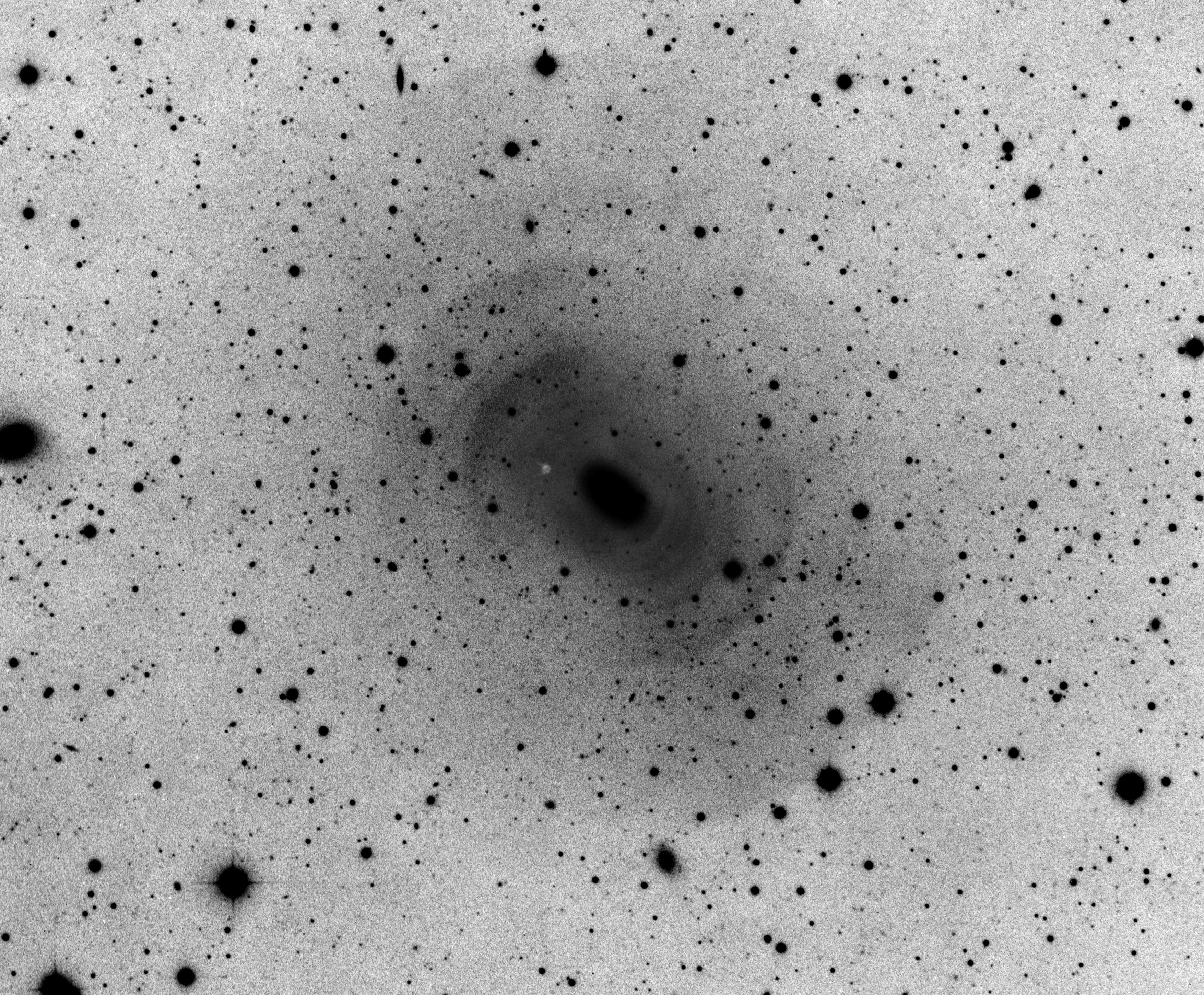}
\caption[NHC\,3923]{The most studied shell galaxy NGC\,3923 (image credit David Malin, Anglo-Australian Observatory). Type~I shell galaxy.}
\label{fig:n3923}
\end{figure}

Stellar shells (also called ripples) observed in some galaxies are defined as arc-like glowing features which have the center of curvature in the center of their host galaxy. See Figs.~\ref{fig:n3923}--\ref{fig:n2764} for examples. The class of shell galaxies was first established by \citet{arpbook} in his ``Atlas of peculiar galaxies'' and in the accompanying paper \citet{arppaper}. The only catalog specialized on shell galaxies was created by \citet{MC83}. It lists 137 object with declination below -17$^\circ$. A~lot of new shell galaxies were discovered later (e.g., \citealp{atkinson13, duc11, ramos11}).

\subsection{Observational characteristics}
\label{sec:obs}
Shells are faint features. Their total luminosity represents typically a~few percent of the total luminosity of the galaxy \citep{DC86}. The brightest shells have the surface brightness of around 24\,mag\,arcsec$^{-2}$ in the V~band \citep{maxSB}. Shells are observed down to the current detection limit (around 30\,mag\,arcsec$^{-2}$).  Based on observations, \citet{atkinson13}  claimed that the majority of tidal features in early-type galaxies (ETGs, i.e. elliptical and lenticular galaxies) occur at the surface brightness of 28\,mag\,arcsec$^{-2}$ or fainter in $V$. The cosmological $\Lambda$CDM simulations  of  \citet{johnston08} predict that most tidal features have surface brightness between 30 and 33\,mag\,arcsec$^{-2}$.

Most galaxies in Malin \& Carter's catalog have less than 4 shells. The galaxy NGC\,3923 holds the record with its 42 shells \citep{bil15b}. Only one shell was detected in some galaxies. The number of the detected shells obviously  depends on the surface-brightness limit of the observation.  

The ratio of the radii of the outermost and the innermost shell in a~galaxy is called the radial range. The record holds NGC\,3923 again, for which it is 108 \citep{bil15b}. 

Shells edges are almost circular. The degree of ellipticity increases with the ellipticity of the host galaxy. The typical ellipticity of shells is around 0.15 in the E3-E4 galaxies, while the E0 galaxies usually have perfectly circular shells \citep{DC86}. 

\begin{figure}[t]
 \centering\includegraphics[width=0.7\textwidth]{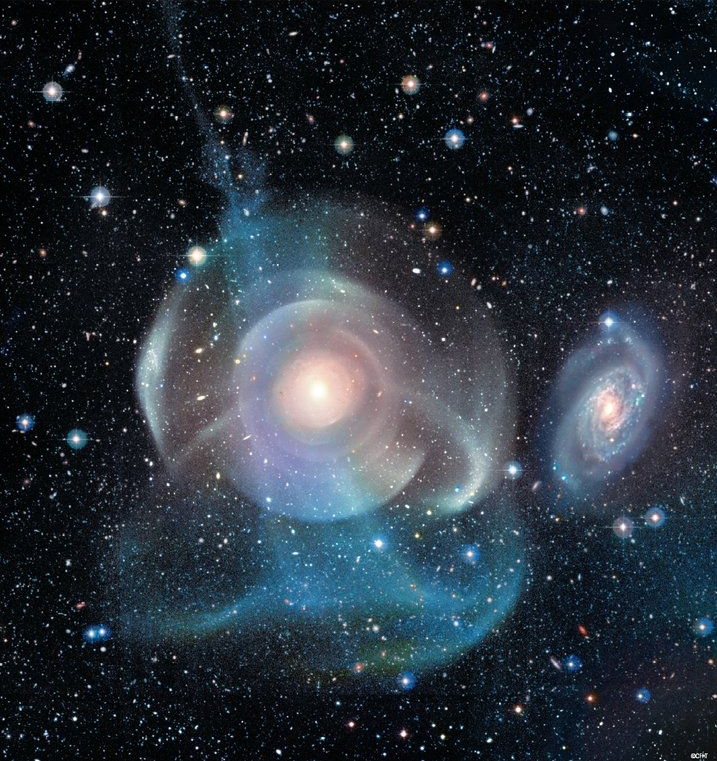}
\caption[NGC\,474]{Shell galaxy NGC\,474 (image credit Pierre-Alain Duc, Canada-France-Hawaii Telescope). Type~II.}
\label{fig:n474}
\end{figure}

The galaxies which appear elongated in projection often have their shells aligned with their major photometric axes (Figs.~\ref{fig:n7600}, \ref{fig:n3923}, \ref{fig:p6510}). Shells are usually randomly distributed around E0 galaxies (Fig.~\ref{fig:n474}) \citep{DC86}.

Shell galaxies are traditionally divided into three morphological types introduced by \citet{prieurdiz} (see also \citealp{prieur90} or \citealp{wilkinson87}).
\begin{itemize}
\item Type I -- The shells form an axially symmetric structure. The symmetry axis coincides with the major photometric axis of the host galaxy. The separation between the neighboring shells increases with radius. The shells tend to be interleaved with radius, i.e. if the shells are sorted with respect to their radius, each shell lies on the opposite side of the galaxy than its immediate predecessor and successor. Examples: NGC\,3923 (\fig{n3923}), NGC\,7600 (\fig{n7600}), PGC\,6510 (\fig{p6510}).

\begin{figure}
 \centering\includegraphics[width=0.65\textwidth]{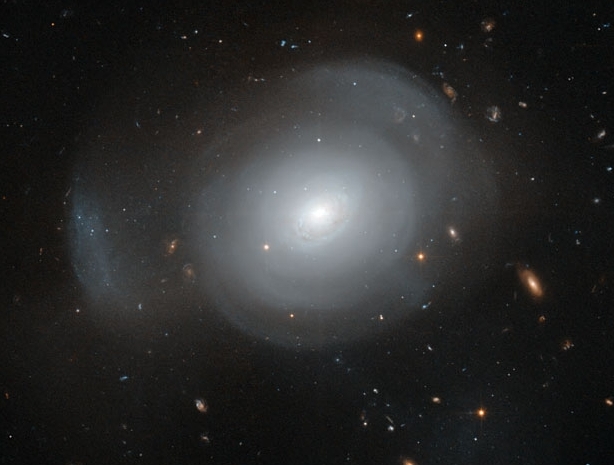}
\caption[PGC\,6240]{Shell galaxy PGC\,6240 (image credit Hubble Space Telescope). Type~II.}
\label{fig:p6240}
\end{figure}

\item Type II -- The shells are randomly distributed in azimuth. Example: NGC\,474 (\fig{n474}).
\item Type III -- The object which do not fit into the previous categories or have only one shell. Examples: NGC\,1316 (\fig{n1316}), NGC\,1344 (\fig{n1344}), NGC\,2764 (\fig{n2764}).
\end{itemize}

There is no special relation between the colors of the shells and their host galaxies. They can be redder, bluer, or the same as the host galaxy \citep{fort86,prieur88, turnbull99,sikkema07, mihos13, gu13}. The color can vary among the individual shells in a~galaxy or even in one particular shell (see the image of NGC\,474 -- \fig{n474}).

Shells are observed in 6\% of all the lenticular galaxies, 10\% of the ellipticals, and around 1\% of the spirals \citep{schweisei88}. Shells occurrence shows environmental dependence. \citet{MC83} note that 47.5\% items of their catalog are isolated galaxies; 30.9\% are in loose groups;  18\% in groups of 2-5 galaxies; and only 3.6\% occur in rich groups or clusters.  There are even indications that a~shell is present in Milky Way \citep{helmi03,deason13}. Shell structures were reported in the Fornax dwarf spheroidal galaxy \citep{coleman04,coleman05}.

\begin{figure}
 \centering\includegraphics[width=0.6\textwidth]{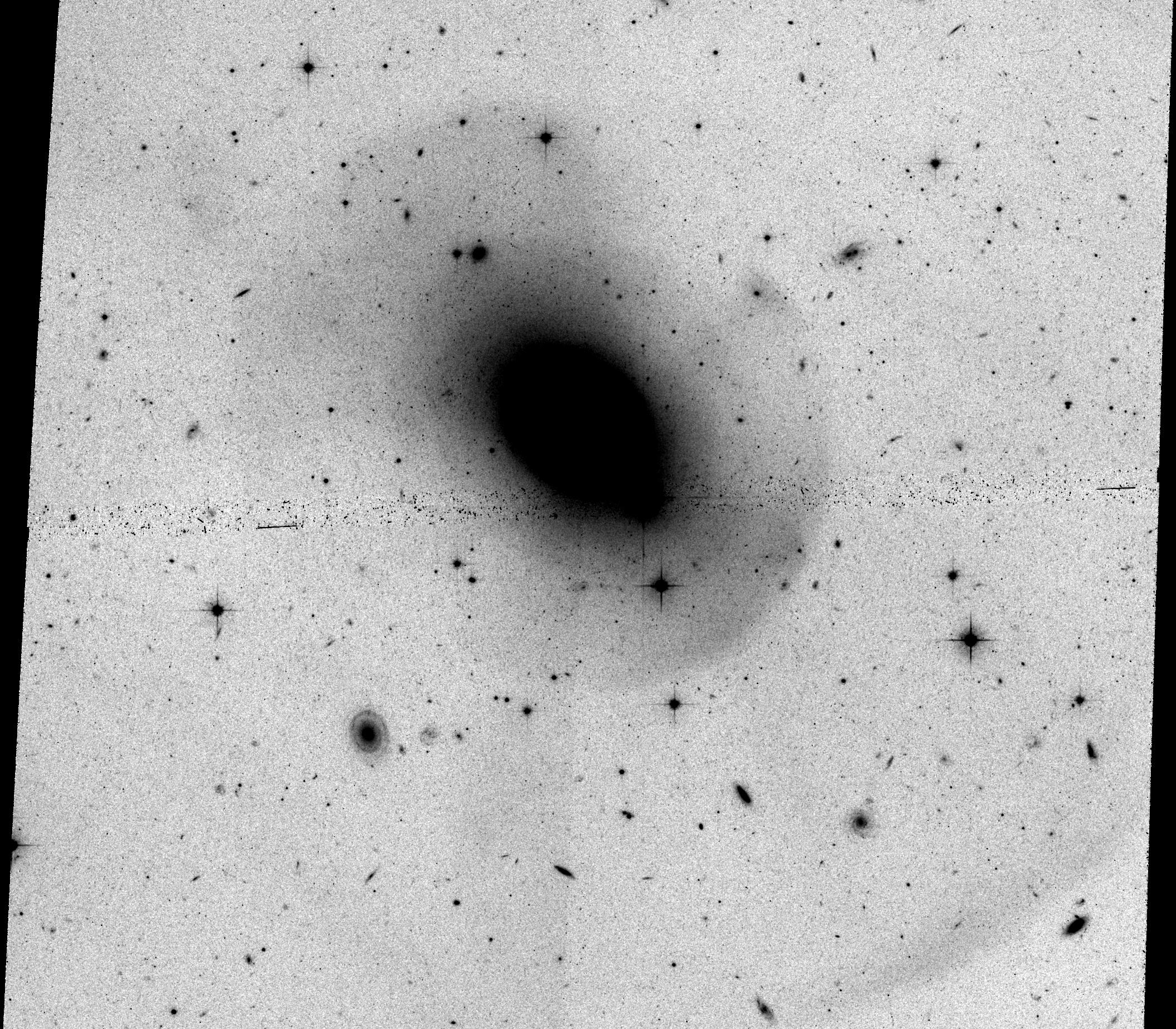}
\caption[PGC\,6510]{Shell galaxy PGC\,6510 (image credit Hubble Space Telescope). Type~I.}
\label{fig:p6510}
\end{figure}

\begin{figure}
 \centering\includegraphics[width=0.6\textwidth]{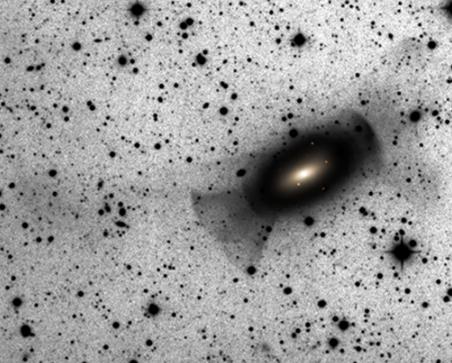}
\caption[NGC\,7600]{Shell galaxy NGC\,7600 (image credit Ken Crawford). Type~I.}
\label{fig:n7600}
\end{figure}

\begin{figure}
 \centering\includegraphics[width=0.7\textwidth]{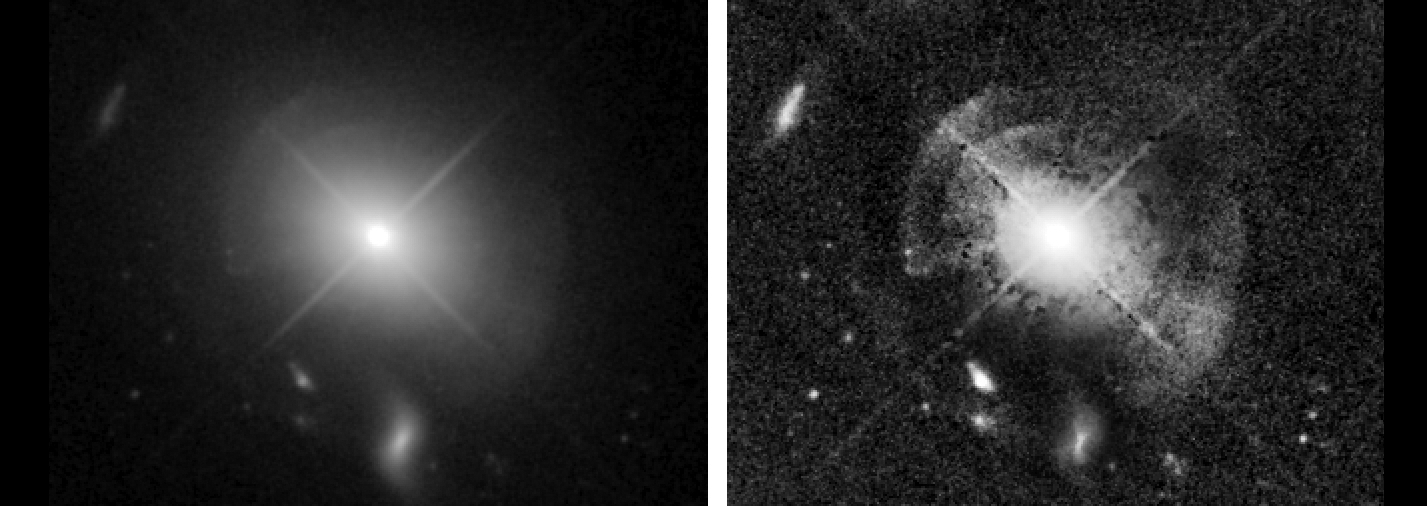}
\caption[MC2\,1635+119 -- Shell galaxy with a quasar]{Galaxy with shells and quasar MC2\,1635+119 (image credit Hubble Space Telescope). Type~I.}
\label{fig:shquasar}
\end{figure}

\begin{figure}
 \centering\includegraphics[width=0.5\textwidth]{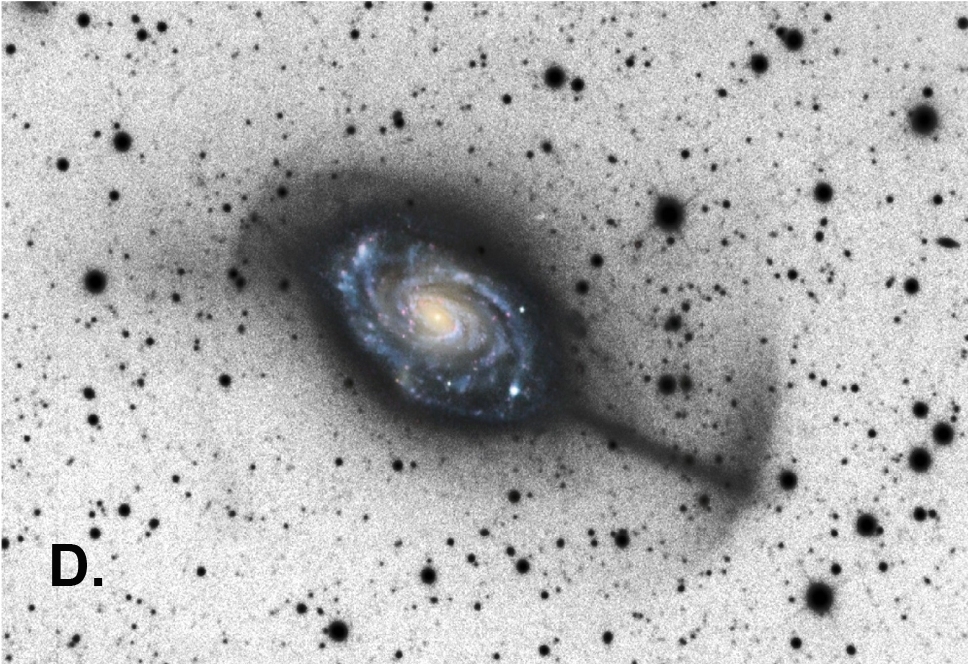}
\caption[NGC\,4651 -- Disk galaxy with a shell]{Rare example of a~disk galaxy with shells -- NGC\,4651, the Umbrella galaxy (image credit Robert Jay GaBany ). Type~I.}
\label{fig:umbrella}
\end{figure}

\begin{figure}
 \centering\includegraphics[width=0.5\textwidth]{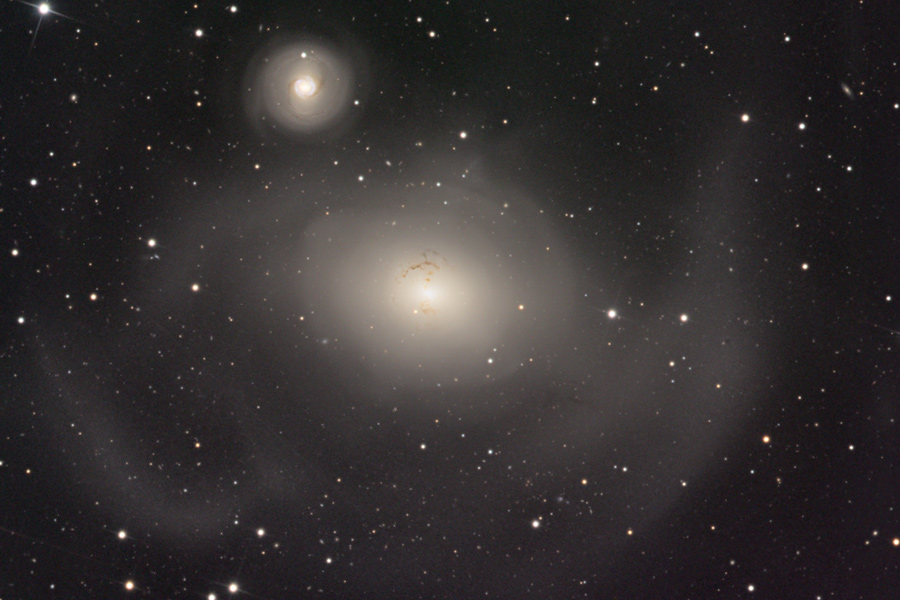}
\caption[NGC\,1316]{Shell galaxy NGC\,1316 (image credit Martin Pugh). Type~III.}
\label{fig:n1316}
\end{figure}

\begin{figure}
 \centering\includegraphics[width=0.5\textwidth]{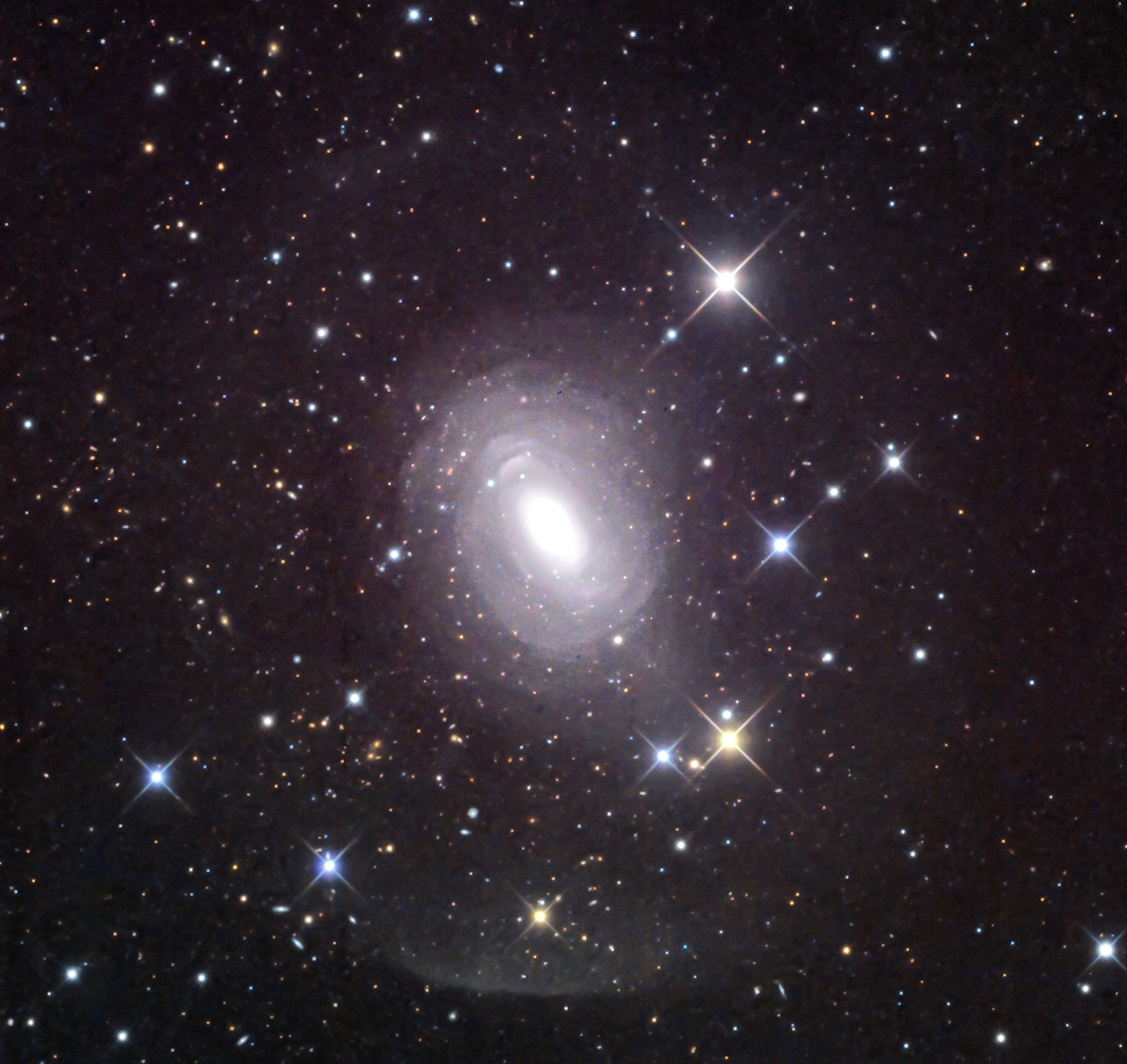}
\caption[NGC\,1344]{Shell galaxy NGC\,1344 (image credit Astrodon Imaging). Originally classified as Type~I, now as Type~III.}
\label{fig:n1344}
\end{figure}

Some shell galaxies contain gas which seems to be correlated with the shells spatially. For example, H\,I and CO clouds were found in Cen~A (NGC\,5128) near the outer edges of the shells \citep{charmandaris, schiminovich94}.  An H\,I emission is seen overlaid over the shells and other tidal features in NGC\,2865 \citep{schiminovich85}. A~loop of H\,I was discovered in NGC\,1210 \citep{hish}. 

\citet{sikkema07} found that shells contain more dust per mass unit than the body of the galaxy in their sample of six objects. They detected dust clouds in the centers of all their objects that are not in the thermal equilibrium, which suggests their external origin.

Shell galaxies often contain signs of a~merger. The shell structure is often accompanied by tidal tails, streams or arms (e.g., \citealp{atkinson13}).  \citet{forbes92} investigated a~sample of 14 ellipticals with kinematically distinct cores. They found that 10 of them had shells.

\subsection{Theories of origin}\label{sec:form}
Several mechanisms of shell formation have been published. Here we list those that are considered feasible today (for more details and a~complete list including historical scenarios see \citealp{ebrovadiz}). Maybe each of them occurred in some galaxy. Now we believe that most shell galaxies were made by the phase-wrapping minor merger model.

\begin{figure}
 \centering\includegraphics[width=0.5\textwidth]{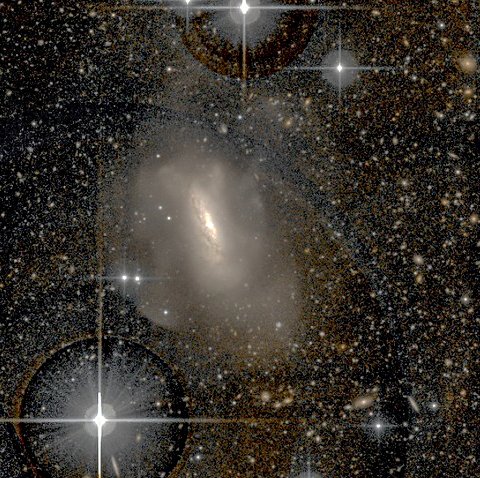}
\caption[NGC\,2764 -- Shells from a major merger]{Shell galaxy NGC\,2764, a~probable result of a~major merger between two disk galaxies (image credit Pierre-Alain Duc, Canada-France Hawaii Telescope). Type~III.}
\label{fig:n2764}
\end{figure}

\subsubsection{Minor merger -- phase wrapping}\label{sec:minmer}
\citet{quinn84} suggested that shells could be remnants of a~small and light galaxy (the secondary) accreted by a~much more massive and bigger galaxy (the primary). When the secondary gets close to the center of the primary, it is disrupted by tidal forces, while the primary is virtually unaffected by the collision. The released stars freely oscillate in the potential well of the primary. When they reach the apocenters of their orbit, they slow down, pile up and form density waves that are observed as the shell edges. There is only a~finite number of shells at a~certain time,  since only the stars which have finished a~half-integer multiple of oscillations are in their apocenter. The shells are formed successively near the primary center and travel outwards. The shell structure can exist for billions of years \citep{DC86}.

 If the encounter has the axial symmetry from the point of view of the observer, then a~Type~I shell system is produced. In the opposite case, a~Type~II or~III shell system is formed. To form and ideal Type~I shell system, we need an exactly radial collision and two elliptical galaxies, or if either of the two is a~disk, then then its symmetry axis must lie in the line of collision. If the collision  is too far from radial, then other kinds of tidal features are created (streams, arms, tails, \ldots, see, e.g., \citealp{johnston08}, \citealp{atkinson13}  or \citealp{amorisco14}). \citet{hq88} were able to produce shells even in a~fly-by of two galaxies where only a~part of the secondary's stars was captured by the primary. 

\begin{figure}
 \centering\includegraphics[width=1\textwidth]{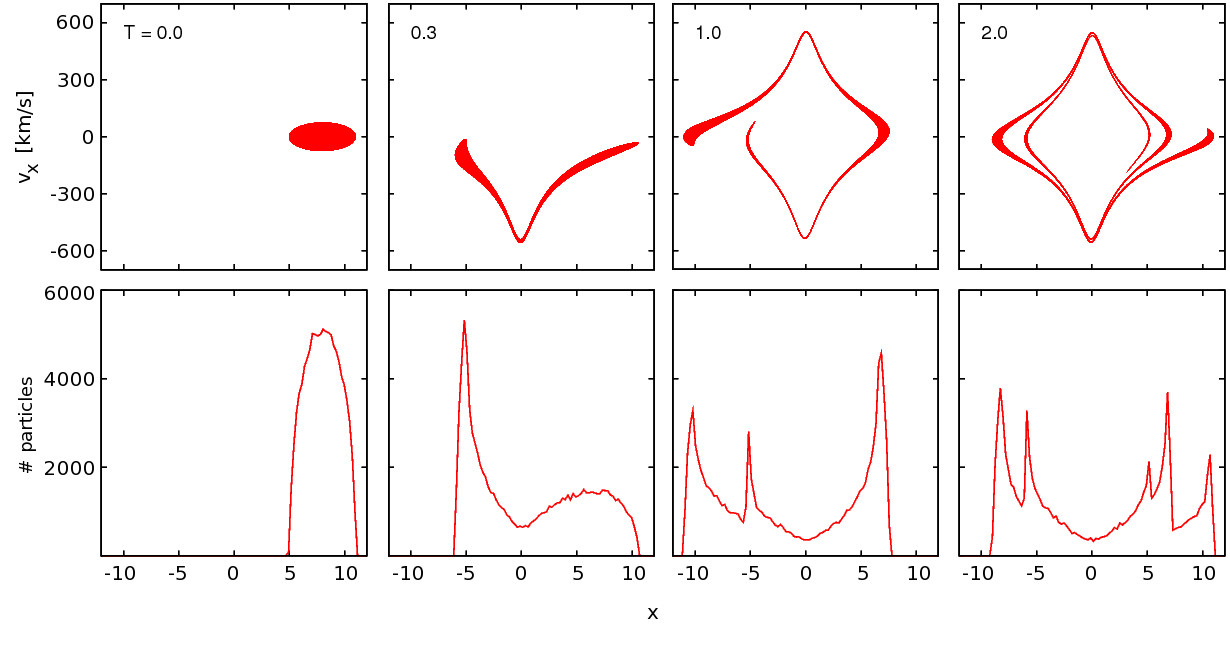}
\caption[Phase-wrapping mechanism]{Illustration of the phase-wrapping mechanism of shell formation in one dimension (image credit Ivana Ebrov{\' a}).  Result of a~restricted three-body simulation (\sect{sim}). Only the particles belonging originally to the secondary are shown. Upper row -- Individual particles velocity vs. coordinate (phase space portrait). Bottom row -- Histogram of particles with respect to their coordinate. Each column corresponds to the time in the top-left corner of the upper images. }
\label{fig:sh1d}
\end{figure}

The name of the model gets apparent from the phase-space portrait (\fig{sh1d}). The secondary forms a~long thin band in the phase space. With the progress of time, the band becomes thinner and longer (the phase-space volume must  conserve because of the Liouville theorem) and  wraps more and more. When this phase-space structure is projected to the coordinate space, sharp-edged features result. The thinner the stripe is, the sharper the edges appear. Thick phase-space folds lead to diffuse shells.

We know that the luminosity of shells represents several percent of the luminosity of the main body of the galaxy. This suggests that the secondary mass has to constitute several percent of the mass of the primary. The sharpest shells are formed from secondaries with a~small phase-space volume, i.e. the secondary has to have a~small size or low velocity dispersion. Given the mass of the secondary, a~disk secondary creates sharper shells than an elliptical being a~dynamically cold system.  For the same reason, we expect that a~collision of two almost equally massive galaxies (say, above the mass ratio of 1:5) does not produce shells, or the shells are very diffuse. However, it comes out that shells can be produced by major mergers under special circumstances, see \sect{majmeree}.

The phase-wrapping model can account for all the observational constraints in most shell galaxies. The only aspect, that has not been fully addressed so far, is the problem of the radial range. Most published simulations use the restricted-three-body simplification (see \sect{sim}) where the secondary is forced to decay in the first pericentric approach to the primary center. The radial range comes out too low in these simulations (see \citealp{DC87} for a~review). \citet{DC87} suggested that the problem could disappear, if the secondary was disrupted gradually: The core of the secondary survives the first pericetric passage and is decelerated by dynamical friction. The released stars form the shells and the core impacts the primary center again giving rise to other shells. The secondary can make several damped oscillations before it dissolves.  The shells formed in the same passage are called a~shell generation \citep{bil13}. Shells from every subsequent generation are located, in average, on lower and lower radii. This way of shell formation was confirmed by self-consistent simulations later \citep{salmon90, segdupii, bartoskovaselfcon, cooper}. Nevertheless, no one of these studies discussed the radial range problem.  The multi-generation origin shell systems also explains the high number of shells observed in some galaxies -- the one-generation simulations show maximally only around 7 shells.

\subsubsection{Minor merger -- space wrapping}
The simulations of \citet{hq89} showed that shells can be produced by the space wraps -- a~sheet of particles wrapped around the galaxy, see \fig{simdisk}. Space wraps are always accompanied by phase wraps in their simulations.  The authors derive theoretically that the space wraps should have different radial surface-brightness profiles. The phase-wrapped shells should have plateau-like surface brightness profile near just under their edge. However, \citet{prieurdiz} claims that the plateau-like profiles were never observed in simulations or in the sky, i.e. the theory might be wrong.

\subsubsection{Major merger of two ellipticals}
\label{sec:majmeree}
The formation of shells in major mergers (mass ratio above around 1:5) of either disk or elliptical galaxies is relatively unexplored. \citet{GGA1,GGA2} presented simulations of collisions of spherical galaxies with and without  DM. In the simulations without DM, they note the formation of shells for a~head-on collision of galaxies with the mass ratio of 1:3. In the simulations with DM, they report shells in the simulations with the mass ratios 1:2 and 1:4. The shells were sharper if the galaxy had a~dark halo. They were made of the stars from the less massive galaxy. The shells resulted from the phase wrapping mechanism like in the minor merger model.

\subsubsection{Major merger of two disk galaxies}
\citet{majorm} suggested that shells could be created in a~major merger of two disk galaxies. They simulated a~prograde collision of two equal disk galaxies where the disk planes lied in the plane of the collision. The body resulting from the merger is an elliptical galaxy surrounded by shells and loops. The authors say that when they plotted the particles in the phase space, they could clearly see 10 wraps. The inner features tend to be aligned with the major axis of the galaxy similarly as observed in the Type~I shell systems. The shells were also created near the galaxy center, in which the simulation was better than the early simulations of the minor mergers. The body of the galaxy followed the de~Vaucouleurs law.

\citet{GGmajordisc} ran a~number of simulations with disk galaxies. Their model contained either bulge, disk and halo, or the bulge was missing. Shells were more easily created in the collisions without the bulge. Shells were formed for all the tested mass ratios -- 1:1, 1:2 and 1:3. They formed Type~II and Type~III shell systems. In contrast with \citet{majorm}, they noted the lack of shells in the prograde mergers of two equal discs. They hypothesized that the perfect spin alignment in the merger of \citet{majorm} may have favored the shell formation. 

Even though the authors do not mention it, shells are created in the simulation\footnote{The video from the simulation can be seen here: \href{https://www.youtube.com/watch?v=FoCglqSStZE}{\url{https://www.youtube.com/watch?v=FoCglqSStZE}}. The shells are formed at 5.4 billion years.} of the future merger of Milky Way with the Andromeda galaxy by \citet{mwm31collision}. A~Type~I shell system is created.  It is visible for about 1\,Gyr. 

In simulations, the shells formed by this mechanism are always accompanied by a~number of other tidal features. The galaxies showing only shells do not probably result from a~major merger involving a~spiral. On the contrary, the observed shell galaxies showing dust lanes, multiple tidal tails and non-relaxed body of the galaxy are considered the results of this process \citep{duc15}, see \fig{n2764}.

\subsubsection{Weak interaction model}
The weak interaction model was proposed in  \citet{wim90} and \citet{wim91}. In this model, shells are density waves in a~thick-disk population of dynamically cold stars induced by a~weak interaction with a~small galaxy which flew near the host galaxy. The model can account for the high shell radial range observed in many galaxies easily. However, this mechanism could not create most shell galaxies: The shells would have to have the same color as the body of the galaxy, which is not generally the case (see \sect{obs}). A~minor axis rotation was detected in some shell galaxies \citep{minrot}, but a~thick disk must rotate along the long axis. \citet{wilkinson00} looked for cold stellar disks in several shell galaxies but they did not found any. This model does not account for the kinematically distinct cores and the increased amount of gas in shell galaxies (see \sect{obs}).

\subsection{Minor merger model -- the results of simulations} 
\label{sec:sim}
The formation of shell galaxies was mostly investigated by the restricted-three-body simulations (see, e.g., \citealp{hq88}). Here, the primary and secondary are modeled as two rigid potential wells which free-fall in the gravitational field of each other. The stars to form the shells are modeled as test particles. At the beginning of the simulations, the test particles move within the  potential well of the secondary. The particles move in the common gravitational field of the two potentials. When the separation between the centers of the potentials is minimal, the potential of the secondary is switched off simulating an abrupt decay. Modifications of this event were also tried: the secondary potential was switched off gradually or when the separation dropped below a~specified value. Then the released particles move freely in the primary potential and form the shells or other tidal features. These simulations produce only one-generation shell systems (the extensions to  multiple generations are possible, see \citealp{segdupii} or \citealp{ebrovadiz}).

\begin{itemize}
\item In a~spherical primary potential, shell edges form almost precise parts of spheres regardless of the type of the secondary \citep{DC86, hq88, hq89}, see Figs.~\ref{fig:simel} and \ref{fig:simdisk}. An axially symmetric collision forms a~Type~I shell system (the axial symmetry, the shells interleaved in radius, the separation of the neighboring shells increases with radius, see \sect{obs}). The shells expand from the host galaxy center. A~shell disappears when it grows too big, so that there is only a~small number of particles that can constitute it. The shell system can exist longer than the Hubble time (\citealp{DC86}, see also Figs.~\ref{fig:simobldisk} and \ref{fig:simproldisk}).

\item In a~spherical host galaxy, the morphology of the shells depends a~lot on whether the secondary is a~disk or elliptical galaxy, as shown by \citet{DC86} and \citet{hq88} (see Figs.~\ref{fig:simel} and \ref{fig:simdisk}). They compared the accretion of a~spherical and a~disk secondary along a~radial trajectory. The secondary-disk axis was tilted by 45$^\circ$ with respect to the collision axis. The accretion of the elliptical secondary created a~Type~I shell system. The result of the merger with the disk  was much more complex. The shells did not have a~common axis. The secondary debris moreover created lobes and bow-tie shapes. The shells, however, still had almost circular edges.

\item If the secondary is elliptical and its mass $m$ is given, then the sharpness of the shells depends on the secondary velocity dispersion (which scales as $\sigma^2\propto m/b$, where $b$ is the half-mass radius). The secondaries with a~higher velocity dispersion lead to more diffuse shells. This is expected, because they have a~larger phase space volume, which scales as $(m/\sigma)^3$. For high-phase-space-volume secondaries, some shells  form complete spheres enclosing the primary center \citep{hq88}.

\begin{figure}
	\centering\includegraphics[width=0.8\textwidth]{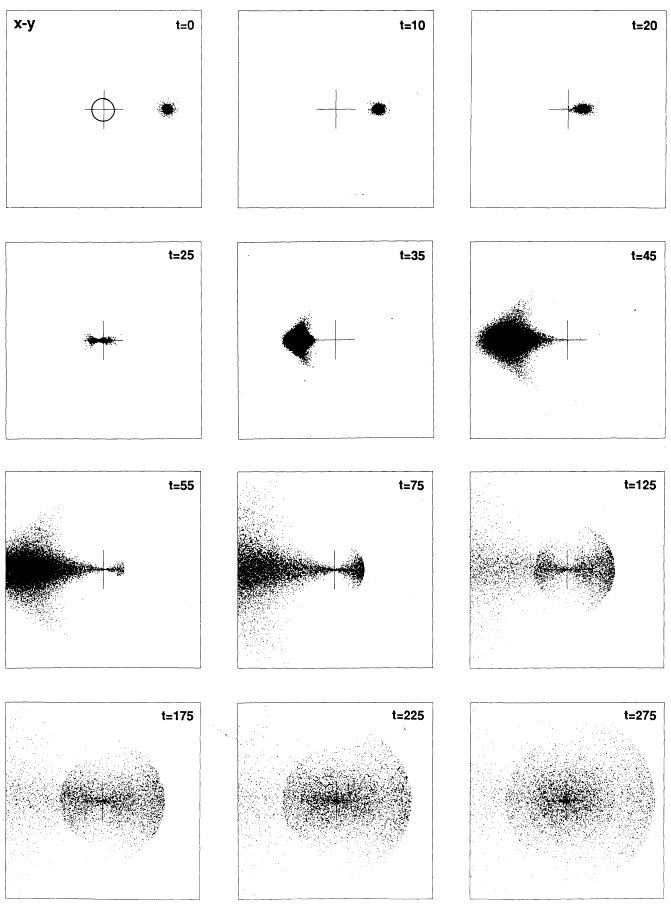}
\caption[Radial accretion of a~spherical secondary on a~spherical primary]{Radial accretion of a~spherical secondary on a~spherical primary (from \citealp{hq88}).  Result of a~restricted-three-body simulation. }
\label{fig:simel}
\end{figure}

\begin{figure}
	\centering\includegraphics[width=0.8\textwidth]{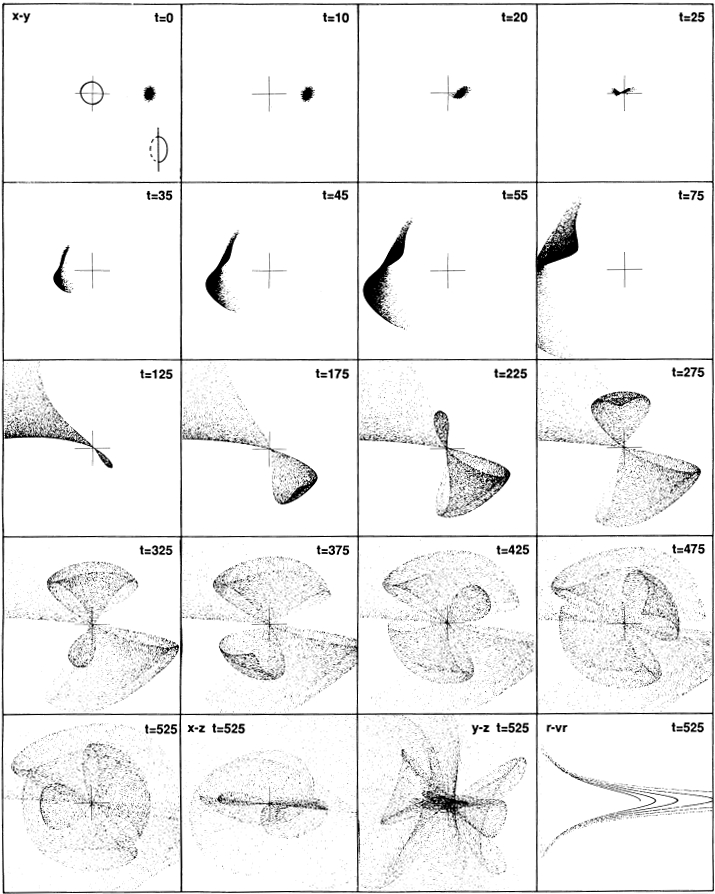}
\caption[Radial accretion of an inclined disk secondary on a~spherical primary]{Radial accretion of an inclined disk secondary on a~spherical primary (from \citealp{hq88}). Some of the shells were formed by the space-wrapping mechanism. Result of a~restricted-three-body simulation. The images in the bottom row show the galaxy from different planes. The last image shows a~projection to the phase space. }
\label{fig:simdisk}
\end{figure}

\item Shells can be even created after a~fly-by of the secondary without a~complete merger. Part of the secondary's particles is captured by the primary. \citet{hq88} simulated this situation with a~disk and a~spherical secondary. In both cases, the captured particles produce sharp-edged features. Most features in the simulation with the disk secondary are evidently non-circular. They seem to be space wraps (the authors do not specify it). In the simulation with the spherical secondary, the features with circular edges seem to be the phase wraps. The centers of the circles do not lie in the center of the primary, though.

\item In a~prolate elliptical potential, the shells tend to be aligned with the symmetry axis, even if the collision is not along the symmetry axis. If the potential is oblate, then the shells form in the symmetry plane and they are randomly distributed in azimuth \citep{DC86,hq89}. The angular extent is the minimal along the direction of the flattening.

\item If the potential is prolate in the center and gets spherical at large radii, then the shells near the center have a~smaller angular extent than those at large radii \citep{hq89}. This and the previous point demonstrates the ``focusing effect'' of elliptic potentials which helps to keep the particles confined near the long axis of the potential, see \fig{prolobl}. For the same reason, shell opening angles are very narrow if a~spherical secondary hits a~disk primary radially in the plane of the disk \citep{hq89}.

\item If the potential is prolate and the secondary is a~disk, then the focusing effect holds the particles near the symmetry axis so the shell system still appears relatively symmetric (Figs.~4 and~14 of \citealp{DC86}, see also our \fig{simproldisk}). Moreover, the shell system exhibits rays of increased surface density starting in the galaxy center and ending at the edge of a~shell.

\item The radii of the shells in a~given potential are the same for an elliptical and disk secondary in a~given time after the secondary disruption (Figs.~13 and~14 of \citealp{DC86}). Our unpublished simulations with many different spherical primaries and secondaries showed that the shell radii depend only on the potential of the primary and the time since the switching-off of the secondary potential for radial collisions. However, the positions of shells at a~given time probably also depend on where the secondary disrupts. 

\item \citet{DC86} found that the shells formed in a~prolate potential have more contrast than in an oblate potential. In projection, we have 60\% probability to see shells in a~prolate potential, but only 20\% in an oblate potential. If the prolate and oblate  elliptical galaxies are represented equally in the Universe, then shell galaxies should be preferably prolate. 

\item The ellipticity of the shells is linked with the ellipticity of the potential (and not with the ellipticity of the mass distribution, \citealp{DC86}).

\begin{figure}[t]
	\centering\includegraphics[width=0.8\textwidth]{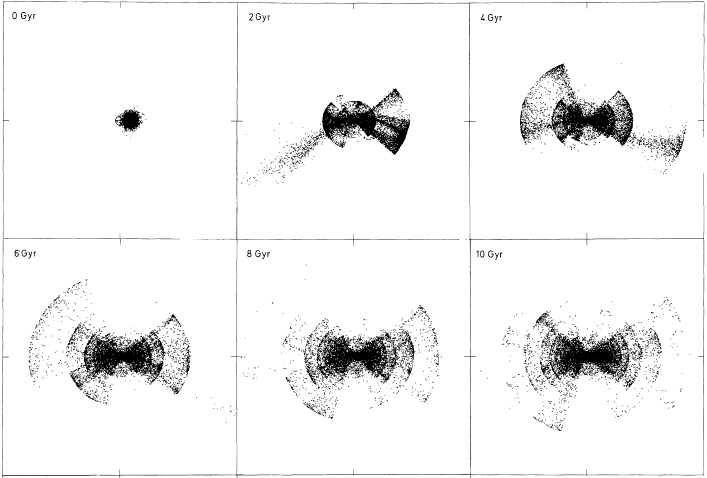}
\caption[Radial accretion of a~disk secondary on a~prolate elliptical primary along the primary major axis ]{Radial accretion of a~disk secondary on a~prolate elliptical primary along the primary major axis (from \citealp{DC86}). The disk is initially seen face-on. The primary is seen edge-on and its major axis is horizontal. Result of a~restricted-three-body simulation.  }
\label{fig:simproldisk}
\end{figure}

\item Our unpublished simulations showed that even a~radial merger can form a~Type~II or III system between spherical galaxies under certain circumstances. The shell system is still axially symmetric but it does not meet the criteria on the Type~I system. For example, the shells can encircle the whole galaxy or the interleaving in radius can be broken. This happens if the orbits of the stars forming the shells show substantial apsidal rotation. In the elliptical potentials, this effect is possibly attenuated by the focusing effect. 
\end{itemize}

A few self-consistent shell simulations have also been published. Here we state their most important results.

\begin{itemize}
\item A~part of the secondary often survives the initial pericentric passage \citep{salmon90, segdupii, bartoskovaselfcon, cooper}.  If the secondary survives, it is decelerated by the dynamical friction, so that it starts making damped oscillations. It can do up to around 4 oscillations before it gets dissolved completely (as we can say from the video\footnote{\label{lab:cooper}The video from the simulation can be seen here: \href{https://vimeo.com/32271838}{https://vimeo.com/32271838}} showing the simulation of \citealp{cooper}).

\begin{figure}
	\centering\includegraphics[width=0.8\textwidth]{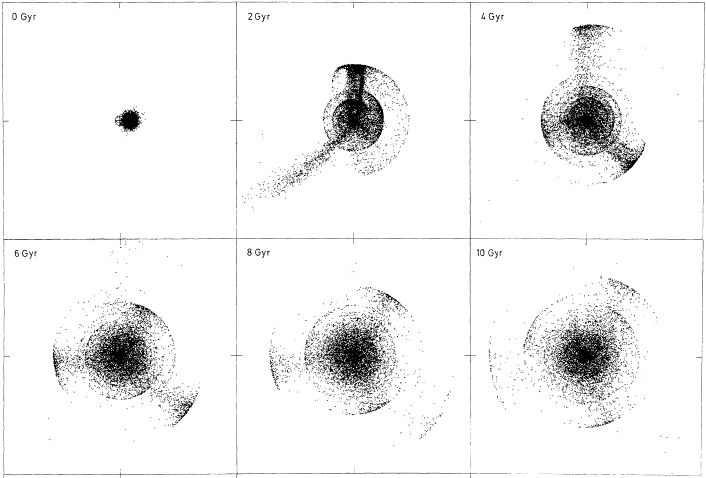}
\caption[Radial accretion of a~disk secondary on a~oblate elliptical primary in the primary symmetry plane ]{Radial accretion of a~disk secondary on a~oblate elliptical primary in the primary symmetry plane (from \citealp{DC86}). The disk is initially seen face-on. The primary is seen face-on. Result of a~restricted-three-body simulation. Note the similarity with NGC\,474 (\fig{n474}). }
\label{fig:simobldisk}
\end{figure}

\begin{figure}
	\centering\includegraphics[width=0.7\textwidth]{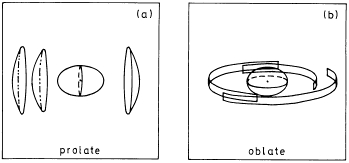}
\caption[Spatial configuration of shells in a~prolate and an oblate primary]{Typical spatial configuration of shells in a~prolate primary (left) and in an oblate primary (right)  (from \citealp{DC86}).  }
\label{fig:prolobl}
\end{figure}

\item The mass loss is the highest in the pericenter (Barto{\v s}kov{\' a}, \textit{in prep.}). 
\item The loss of the orbital energy is the highest in the pericenter  \citep{segdupii}.
\item Each further shell generation is located, in average, at lower galactocentric radii than the previous one \citep{salmon90, segdupii, bartoskovaselfcon, cooper}. It probably solves the radial range problem.
\item The rate of tidal disruption is highly dependent on the impact parameter \citep{segdupii}.
\item Let  $\rho_\mathrm{sec}$ be the central density of the secondary, $\rho_\mathrm{prim}$ the density of the primary at the pericentric radius of the secondary and $\eta = \rho_\mathrm{prim}/\rho_\mathrm{sec}$. If $\eta \approx \rho_\mathrm{sec}^{-1}$, then the secondary dissolves very easily. More compact secondaries survive better \citep{segdupii}.
\item If $\eta \approx \rho_\mathrm{prim}$, then tidal striping depends on the radiality of the collision: secondaries on the more radial orbits are affected more \citep{segdupii}.
\item The magnitude of dynamical friction is little sensitive to the impact parameter \citep{segdupii}.
\item Dynamical friction depends by far most on the concentration of the secondary: it is lower for diffuse secondaries. The energy loss also depends, but not so strongly, on the density profile of the secondary \citep{segdupii}.
\item Dynamical fiction merely depends on $\rho_\mathrm{sec}$ \citep{segdupii}.
\item We could expect that the formation of Type~I shell galaxies is extremely improbable in the minor merger scenario, because the collision has to have the axial symmetry. However, we can see around nine Type~I shell galaxies in the cosmological simulation of \citet{cooper} (see the video linked in footnote~\ref{lab:cooper}).
\end{itemize}

\subsection{Analytic modeling of Type~I shell galaxies}
\subsubsection{Shell radii}\label{sec:shrad}
The time evolution of shell radii of Type~I shell systems can be modeled analytically. This can be used for constraining the gravitational potential of shell galaxies (see \sect{constr}). Several precise and approximate formulas appeared in literature. We describe them in this section. We assume the formation of shells by the phase-wrapping minor merger model (see \sect{minmer}).

Type~I shell systems are formed by radial collisions. Thus, to derive the analytic equations let us assume that the stars forming the shells move on exactly radial orbits and we will consider only the particles moving along the major axis. We further assume that the gravitational potential $\phi(r)$ is symmetric and that the stars are released from the secondary when it goes through the primary center.  Each shell is characterized by the serial number $n = 0, 1, 2,\ldots$ It denotes the number of pericentric passages made by the stars constituting the given shell after their releasing from the secondary. The initial passage, when the stars are released from the secondary, does not count in. The radius of a~shell is measured from the primary center to the edge of the shell.

The simplest formula by \citet{quinn84} comes from the fact that shell edges are made of stars close to their apocenters. If $t_\mathrm{A}$ is the time since the releasing of the stars from the secondary, then the $n$-th shell is  located approximately at the radius $r_{\mathrm{A},n}$ meeting the condition
\begin{equation}
t_\mathrm{A} = \left(n+1/2\right) P(r_{\mathrm{A},n}),
\label{eq:hqform}
\end{equation}
where $P(r)$ denotes the time needed for a~particle with apocenter at the radius $r$ to oscillate between two subsequent apocenters. In other words, $P(r)$ is twice the free fall time at the radius $r$. We can easily derive that
\begin{equation}
	P(r) = \sqrt{2}\int_0^r \left[\phi(r)-\phi(x)\right]^{-1/2}\d x.
	\label{eq:pr}
\end{equation}
For the sake of simplicity, let us call the radius $r_{\mathrm{A},n}$ the apocentric radius of the $n$-th shell.

But the actual radius of a~shell, $r_n$ actually differs a~little from the point where the stars are just in the apocenter, i.e. $r_{\mathrm{A},n}$, because the shell moves \citep{DC86}. To see this, imagine the surface enveloping the edge of the shell and follows it in time. In order for the particles not to cross the surface, the surface must move at least with the same radial velocity of the particles that are just reaching it. Similarly, in order for the surface not to recede from the particles, it must move at most with the same velocity as the particles that are just reaching it. Altogether, the particles at the edge of a~shell move with the phase velocity of the edge. We can easily see that in given time, the actual radius of a~shell $r_n$ is greater than $r_{\mathrm{A},n}$ resulting from the equation \equ{hqform}.  Simulations show that the apocentric and the shell radius of the first shell differ by around 10-20\%. The difference is lower for higher serial numbers. This follows from the fact the velocity of a~shell decreases with the increasing serial number.

Time derivative of \equ{hqform} implies that the velocity of the $n$-th shell is
\begin{equation}
v_n(r) \approx v_{\mathrm{A},n}(r) =  \frac{1}{\left(n+1/2\right)\frac{\d P(r)}{\d r}},
\label{eq:vs}
\end{equation}
where $v_n$ is the actual velocity of the $n$-th shell and $v_{\mathrm{A},n} = \frac{\d r_{\mathrm{A},n}}{\d t_\mathrm{A}}$ is the velocity of the apocentric radius. We can see that the velocity of a~shell at a~given radius decreases with the increasing serial number.

Let us assume that the gravitational field is approximately homogeneous between $r_{\mathrm{A},n}$ and $r_n$ and its magnitude is $a(r_n)$. When the particles are at the radius $r_n$ at the time $t_\mathrm{E}$, they have the velocity of $v_n$. They are decelerated by the gravitational acceleration of $a(r_n)$, until they reach zero radial velocity at the radius $r_{\mathrm{A},n}$. The time needed for a~particle to get from $r_{n}$ to $r_{\mathrm{A},n}$  must meet
\begin{equation}
0 = v_n(t_\mathrm{E}) -(t_\mathrm{A}-t_\mathrm{E})a(r_n).
\end{equation}
Now we use the approximation
\begin{equation}
	v_n(t_\mathrm{E})\approx v_n(t_\mathrm{A}) \approx v_{\mathrm{A},n}(t_\mathrm{A})
	\label{eq:vapprox}
\end{equation}
assuming first that the velocity of shell edge varies slowly in time and then that $v_{\mathrm{A},n}(t)$ is almost $v_n(t)$ since $r_{\mathrm{A},n}$ and $r_n$ are always very close to each other (we know this from our unpublished restricted-three-body simulations).  Finally we get
\begin{equation}
	t_\mathrm{A}-t_\mathrm{E} \approx \frac{v_{\mathrm{A},n}(t_\mathrm{A})}{a(r_n)},
	\label{eq:tate}
\end{equation}
where $a$ is taken as positive. The separation between $r_{\mathrm{A},n}$ and $r_n$ is therefore
\begin{equation}
r_n(t_\mathrm{E}) - r_{\mathrm{A},n}(t_\mathrm{A}) = \frac{1}{2}a(r_n)(t_\mathrm{A}-t_\mathrm{E})^2 \approx \frac{1}{2}\frac{v_{\mathrm{A},n}^2(t_\mathrm{A})}{a(r_n)}.
\end{equation}
The last two equations allow us to correct the result of Eqs.~\ref{eq:hqform} and~\ref{eq:pr} to get a~better approximation of the shell radii  \citep{ebrova12}.

We can do another kind of approximation of a~similar nature \citep{bil13}. It removes the necessity to approximate the gravitational field between $r_{\mathrm{A},n}$ and $r_n$ by the homogeneous field. We start form the conservation of energy
\begin{equation}
	\frac{1}{2}v_n^2(t_\mathrm{E}) = \phi\left[r_{\mathrm{A},n}(t_\mathrm{A})\right]-\phi\left[r_n(t_\mathrm{E})\right].
\end{equation}  
Therefore
\begin{equation}
	r_n(t_\mathrm{E}) = \phi^{-1}\left\{\phi\left[r_{\mathrm{A},n}(t_\mathrm{A})\right]-\frac{1}{2}v_n^2(t_\mathrm{E})\right\} \approx \phi^{-1}\left\{\phi\left[r_{\mathrm{A},n}(t_\mathrm{A})\right]-\frac{1}{2}v_{\mathrm{A},n}^2(t_\mathrm{A})\right\},
	\label{eq:rn}
\end{equation}
where we used \equ{vapprox} again. Now we can correct the results of Eqs.~\ref{eq:hqform} and~\ref{eq:pr} by Eqs.~\ref{eq:rn} and~\ref{eq:tate} to get a~better approximation of shell radii. The shell radii calculated by this method differ by around 1\% from the precise formulas for any time after the merger (B{\' i}lek et al., \textit{in preparation}). 

The first precise method for calculating shell radii at a~given time appeared in \citet{ebrova12}. It is based on the equation
\begin{equation}
\begin{array}{rcl}
t_\mathrm{E}=n\sqrt{2} & \int_{0}^{r_{\mathrm{A},n}} & \left[\phi(r_{\mathrm{A},n})-\phi(x)\right]^{-1/2}\d x+\\
+ \frac{1}{\sqrt{2}}& \int_{0}^{r_\mathrm{*,n}} & \left[\phi(r_{\mathrm{A},n})-\phi(x)\right]^{-1/2}\d x.
\end{array}
\end{equation}
First, we choose the time when we want to calculate the radius of the shell, $t_\mathrm{E}$, and its serial number, $n$. If $r_{\mathrm{A},n}$ is given in addition, then we can determine the value of $r_{\mathrm{*,n}}$ so that the equality holds true. The shell edge radius $r_n$ is the maximum of $r_\mathrm{*,n}$ over all $r_{\mathrm{A},n}$. The radius $r_n$ has to be found iteratively.

Another precise algorithm will be derived in B{\' i}lek et al. (\textit{in preparation}). It is based on the formulas
\begin{equation}
	\frac{\sqrt{8}n P'(r_{\mathrm{A},n})}{\phi'(r_{\mathrm{A},n})} = \int_0^{r_n}\left[\phi(r_{\mathrm{A},n})-\phi(x)\right]^{-3/2}\d x 
	\label{eq:bilprep1}
\end{equation}
and
\begin{equation}
	t_\mathrm{E} = nP(r_{\mathrm{A},n})+\frac{1}{\sqrt{2}}\int_0^{r_n}\left[\phi(r_{\mathrm{A},n})-\phi(x)\right]^{-1/2}\d x.
	\label{eq:bilprep2}
\end{equation}
The function $P(x)$  is defined by \equ{pr} and $P'(x)$ is its derivative with respect to $x$. We choose a~value of $r_{\mathrm{A},n}$. Then we evaluate the left-hand side of \equ{bilprep1}, $L$, for the serial number of the shell of interest, $n$. Then we integrate the integrand of the right-hand side from zero until we reach the value of $L$. This integration bound is $r_n$ we look for. The time when the shell has the radius of $r_n$ can be calculated from \equ{bilprep2}. If we repeat this procedure for many values of $r_{\mathrm{A},n}$, we get the time evolution of the radius of the $n$-th shell. These precise formulas have never been compared to simulations (it will be done in B{\' i}lek et al., \textit{in preparation}).

In a~Type~I shell galaxy, the shells with odd serial numbers lie on the opposite side of the galaxy than those with even serial numbers. For this reason, it is advantageous to add a~sign to the shell radii in our model. We will treat the radii of the shells with odd serial numbers as positive and as negative for the even serial numbers.

Next, we have to account for the multiple shell generations present in the system (see \sect{minmer}). Let $t_N$, $N=1,2,\ldots, N_\mathrm{max}$, be the ages of the individual generations (the times since their formation at the moment of the observation). We calculate the shell radii, $r_n(t_N)$, for each generation using one of the above-mentioned methods. At each pericentric passage, the secondary hits the center of the primary from the opposite side than at the previous passage. To account for this, we multiply the shell radii belonging to the generations with odd numbers by -1. The set of shell radii in our modeled system is described as
\begin{equation}
	\left\{(-1)^{N+n+1}r_n(t_N); n=0,1,\ldots,n_{\mathrm{max},N}; N=1,2,\ldots,N_\mathrm{max} \right\}
\end{equation}
The ages $t_N$ depend mostly on the initial infall velocity of the secondary and dynamical friction during individual pericentric passages. So far, no attempt to calculate them analytically has been done. It may be expected that the differences $t_{N+1}-t_N$ decrease with increasing generation number $N$ since the secondary oscillations are damped.

However, in real galaxies the shells form \textit{gradually} and they disappear. Let us assume that the stars released during every particular pericentric passage have a~one-peaked radial-energy distribution; this was confirmed by our restricted-three-body simulations (however, the surviving core of the secondary could possibly break this assumption). In a~given time, the shells with low serial numbers could disappear and the shells with high serial numbers could have not been formed yet. In total, the set of the radii of the observable shells can be expressed as 
\begin{equation}
	\left\{(-1)^{N+n+1}r_n(t_N); n=n_{\mathrm{min},N},n_{\mathrm{min},N}+1,n_{\mathrm{min},N}+2,\ldots,n_{\mathrm{max},N}; N=1,2,\ldots,N_\mathrm{max} \right\}.
	\label{eq:allsh}
\end{equation}

\subsubsection{Azimuthal surface-brightness profiles}
The angular surface-brightness profile of shells depends on several factors. As the simulations of \citep{hq88} showed, it depends on the original size and velocity dispersion of the secondary. Then the shape of the primary potential is also important: The rate of the apsidal precession of orbits depends on the central concentration of the potential. It can be expected, and our unpublished simulations confirm it, that the shell opening angle increases with increasing serial number in a~spherical potential because of the apsidal precession. Moreover the shell opening angle is highly affected by the focusing effect of elliptical potentials, see \sect{sim}. The impact parameter probably also influences the shell azimuthal profile.

\section{Testing MOND in shell galaxies}\label{sec:mondsh}
MOND has been tested mainly in disk galaxies so far. They usually contain atomic hydrogen allowing us to trace the kinematics up to large radii. We can often measure the rotational curves down to the acceleration of $a_0/10$ \citep{famaey12}. It is difficult to test MOND in elliptical galaxies because they lack kinematic tracers on known orbits in the regions where $a \ll a_0$ (see \citealp{milg12} for details). These are the classical ways of investigating the gravitational field in elliptical galaxies:  
\begin{description}
	\item[Jeans analysis] -- The Jeans equations imply the relation between the mass profile and the velocity-dispersion profile. This can be used for constraining the gravitational potential by tracers like stars, globular clusters, planetary nebulae and, after the galaxy stacking, also satellites.  The weakest acceleration field probed by this method in individual galaxies reaches about $a_0/3$ (e.g., \citealp{samur14}). The Jeans analysis does not allow a~precise testing of the MOND equations because of the anisotropy degeneracy. 
	\item[X-ray gas] -- Most elliptical galaxies contain hot X-ray emitting gas. Supposing the equilibrium between the gravity and gas pressure, we can derive the profile of the gravitational field from the temperature profile of the gas. Unfortunately, the temperature profiles  can be usually  measured only near the center of the galaxy. \citet{milg12} used this method to verify MOND down to $a_0/10$ in two ellipticals with an exceptionally high amount of hot gas.
	\item[Gravitational lensing] -- The strong lensing cannot be used for testing MOND because it requires strong gravitational fields on the galactic scales. The weak gravitational lensing cannot be used for individual galaxies (but after stacking a~lot of galaxies, the weak lensing by elliptical galaxies is consistent with MOND, \citealp{milg13}).
	\item[Rotating gas] -- Occasionally, elliptical galaxies contain rotating disks of gas. For example, it was found that the gas rotation curve in the elliptical galaxy  NGC\,2974 agrees with MOND \citep{weij08}.
\end{description}
The low number of tests of MOND in ellipticals down to weak fields raises the question of whether MOND really stems from a~a general law of nature, or it is only a~correlation between the baryonic and dark matter valid for some galaxy types \citep{milg12}.

Shells are very interesting structures for MOND, especially the Type~I systems. They are made of stars in known (radial) orbits and they extend down to low accelerations, often under $a_0/5$. There are simple relations between the gravitational potential and the shell radii and motions (see \sect{shrad}). MOND was originally inspired by disk galaxies where all the stars move on approximately circular orbits. The magnitude of the gravitational acceleration does not vary much in time. On the other hand, the stars constituting the shells travel radially from the hight- to the low-acceleration regions. The investigation of shells could therefore prefer or eliminate some of the modified inertia versions of MOND. If it proved that \equ{algrel}, which works well for the circular orbits, does not work for the radial orbits, it would be an evidence that the correct MOND theory is a~theory of modified inertia (elliptical galaxies are approximately spherical objects). On the contrary, if the equation works well even for radial orbits, it would falsify the theories predicting the opposite. 

Moreover, the process of the secondary accretion is expected to be different for the MOND theories and the DMF. There are two major differences: 1) Dynamical friction is usually weaker in MOND than in the DMF (see \sect{mondsim}); and 2) the decay of the secondary can be faster in MOND because of the EFE (see \sect{impl}), especially if the satellite is diffuse (is in the deep-MOND regime) before its first approach. In MOND, if the secondary is concentrated enough to be immune to the EFE and to resist the tidal stripping (see \sect{sim}), then it probably makes many more oscillations and produces many more shell generations than in the DMF because of the weak dynamical friction. This would have effect on the number of shells, their surface-brightness statistic, radial distribution and the number of the observable surviving secondaries. This is to be explored by self-consistent simulations.

\subsection{Constraining gravitational potential of Type~I shell galaxies from shell radial distribution}\label{sec:constr}
In \sect{shrad} we saw that the evolution of shell radii in Type~I shell galaxies can be modeled analytically once the host galaxy's potential is given. This fact was the inspiration for several methods for constraining the gravitational potential from the shell radial distribution.

The first attempts were done in the 80s \citep{quinn84, DC86, hq87, hqmond, prieur88}. It was shown that elliptical galaxies show mass discrepancy \citep{prieur88,hq87}.  \citet{hq87} attemped to determine the serial numbers of some shells and to deduce the age of the shell system in several galaxies.  \citet{hqmond}  claimed that MOND is inconsistent with the shell distribution in NGC\,3923. The lastly mentioned work was criticized by \citet{milgsh}. Apart from a~few logical errors in the paper, Milgrom pointed to several general downsides of the methods used that time. Most notably, the methods assumed that all the shells came from one generation and that no shells escaped observations. Later on, \citet{DC87} showed the importance of the multiple-generation scenario. These difficulties probably led to fading of the interest in shell galaxies by the end of the 80s.

\subsubsection{Method of shell identification}\label{sec:shid}
In \citet{bil13}, we developed the ``shell identification method'' for testing the consistency of a~given gravitational potential with the observed shell radii in the Type~I shell galaxies. In contrast with the previous methods, it accounts for the multiple shell generations present in the system and for the shells escaping observations. It is based on \equ{allsh} and its assumptions.

Here is how the method works.
\begin{enumerate}
	\item We denote all the observed shells by a~label $\lambda$. Let $R_\lambda$ be the radius of the shell $\lambda$. We add signs to these radii. We divide the image of the galaxy into two halves by a~line going through the center of the galaxy perpendicularly to the symmetry axis of the shell system. We choose one side of the galaxy as reference and mark the  radii of the shells on this side as positive.  The radii of the shells on the opposite side are then negative. 

	\item We calculate the evolution of shell radii in the tested potential. 

	\item The next step is to assign a~shell number $n_\lambda$ and a~generation number $N_\lambda$ to each observed shell $\lambda$. This is what we call the shell identification. However, the identification must satisfy several criteria to be acceptable: 
\begin{itemize}
    \item Generation ages $t_N$, $N = 1, 2, \ldots, N_\mathrm{max}$, must exist, so that the observed shell positions are approximately equal to those calculated at $t_{N_\lambda}$, i.e. $R_\lambda \approx o_{I}(-1)^{N_\lambda+n_\lambda+1}r_{n_\lambda}(t_{N_\lambda})$, for every $\lambda$. The variable $o_I$ is called the sign of the first generation. It takes the value of $+1$ if the shells with the odd serial numbers classified into the first generation lie on the reference side of the galaxy and $-1$ in the opposite case (the shells with the odd and the even serial numbers have to lie on the opposite sides because it is an assumption of \equ{allsh}). The sign of the first generation indicates the side of the galaxy from which the secondary originally flew in.
    \item An acceptable identification should not require a~lot of missing shells. This means for every generation, that the identified shell serial numbers should form a~continuous series of integers (e.g., 3, 4, 5, 6). If a~number is missing in this series, the corresponding shell must be escaping observations. Such shells are called the missing shells. If a~sufficient number of missing shells is allowed, then it is easy to meet the remaining criteria for almost any combination of shell radii and potential. 
    \item Not too many generations are allowed for the same reason.
    \item The $N$-th generation must be older than the $(N+1)$-th generation, for every~$N$.
    \item The differences between ages of subsequent generations are becoming shorter because the amplitude of oscillations of the secondary is damped by dynamical friction. 
\end{itemize}
\end{enumerate}
If such an identification exists, then the tested potential is considered compatible with the observed shell radii. There can be more than one acceptable identification. We see that the identification criteria are somewhat arbitrary. We have to choose them according to our experience from observations and simulations.

If we assume that the tested potential is correct, the method allows us to reveal
\begin{itemize}
	\item The age of the shell system. It is important for the research of galaxy formation.
	\item The ages of the individual generations. Their differences yield information about the dynamical friction during the merger. The friction is known to be weaker in (at least some) MOND theories than in the DMF (see \sect{mondsim}). If we combine the shell identification method with self-consistent simulations, we could possibly distinguish between MOND and the DMF.
	\item Since we know the ages of the generations, we are able to calculate the positions of the shells that have not been discovered so far. 
\end{itemize}

In all the papers on shell identification so far \citep{bil13, bil14, bilcjp}, the method had the following assumptions:
\begin{itemize}
	\item The shells were created by the minor merger model.
	\item The shell system originates from one secondary. This assumption is motivated by the fact that only $Q = \frac{1}{30}$ of all elliptical galaxies contain a~Type~I shell system (see \sect{obs}), i.e. the formation of such systems is rare. The following calculation reveals more details.  Let us assume that the secondaries are accreted independently of each other on a~given primary. Let us denote by $p$ the probability that a~randomly chosen elliptical galaxy accretes \textit{one} secondary that creates a~Type~I shell system observable today. The probability of the accretion of $n$ secondaries on this galaxy is then $p^n$. We have the equation $p+p^2+p^3+\ldots = Q$. Thus, $p = \frac{Q}{1+Q} \approx 3\%$.  From the formula for conditional probability, we obtain that the probability that the shells in a~randomly chosen Type~I shell galaxy were created by $n$ secondaries is $P = \frac{p^n}{Q}  = \frac{30}{31^n}$. Hence the probability that the shells in a~randomly chosen Type~I shell galaxy were created from one secondary is 97\%. 
	\item The merger was exactly radial.
	\item The evolution of shell radii is well described by Eqs.~\ref{eq:hqform}, \ref{eq:tate} and \ref{eq:rn}. This includes the assumption that the stars release from the secondary when it goes through the primary center.
\end{itemize}

\subsection{Consistency of shell radii in NGC\texorpdfstring{\,3923}{Lg} with MOND}\label{sec:bil13}
In the paper \citet{bil13} (B13 hereafter), we tested the consistency of the shell radii in NGC\,3923 with MOND (see the paper attached in \app{bil13} for details). This object is a~well-studied nearby elliptical galaxy lying around 23\,Mpc from us. This corresponds to the linear scale of around 1\,kpc per 10\arcsec{}. It is an exceptional object for several reasons: 1) It hosts the biggest known shell with the radius of around 120\,kpc. 2) Its shell system has the greatest known radial range (see \sect{obs}), which is 65. 3) It has the richest shell system. At the time when B13 was published, 27 shells were known.  

We used the shell radii published in \citet{sikkema07} and \citet{prieur88}. The radii from \citet{sikkema07} were measured from an image taken by the Hubble Space Telescope and cover the inner 130$^{\prime\prime}$ of the galaxy. The radii of the larger shells were adopted from \citet{prieur88}. 

\begin{figure}[ht]
 \centering\includegraphics[width=1\textwidth]{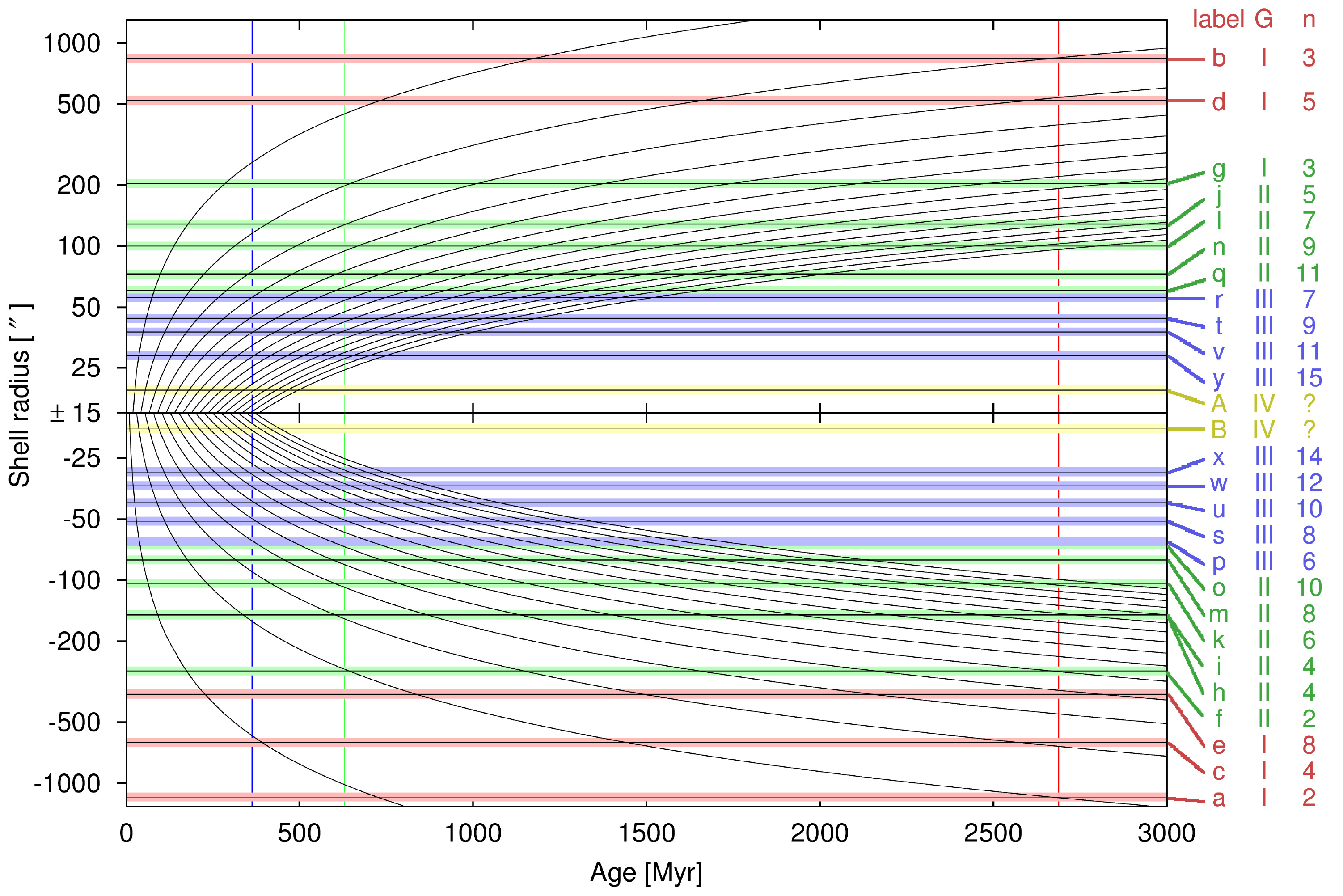}
\caption[Test of the compatibility of the shell radii in NGC\,3923 with MOND]{Test of the compatibility of the shell radii in NGC\,3923 with MOND \citep{bil13}. Horizontal lines -- Observed shell radii multiplied by the corresponding sign (see \sect{shid}). The width of the lines is 5\%. Curved lines -- Calculated evolution of shell radii in the tested potential. Vertical lines -- Times where the observed radii are reproduced by the model. Shells corresponding to the same generation have the same color. Letters -- Designation of the shell according to \tab{sh}. Roman numbers -- Identified generation number of the shell. Arabic numbers -- Identified serial number of the shell.}
\label{fig:shcomp}
\end{figure}

\begin{table}[ht!]
\centering
\begin{tabular}{lrrrccrc}
\hline\hline
$\lambda$ & $R$ [$^{\prime\prime}$] & $R$ $[$kpc$]$ &  $a$ [$a_0$] & $G$ & $n$ & $r$ $[^{\prime\prime}]$ & $\Delta$ [\%] \\
\hline
$a$  &  $+1170$    &  $130.7$ &    $0.2$    & I   & $2$   & $+1178$  & $0.7$ \\
$b$  &  $-840$     &  $93.8$  &    $0.3$    & I   & $3$   & $-845$   & $0.6$ \\
$c$  &  $+630$     &  $70.4$  &    $0.4$    & I   & $4$   & $+658$   & $4.5$ \\
$d$  &  $-520$     &  $58.1$  &    $0.4$    & I   & $5$   & $-539$   & $3.6$ \\
$e$  &  $+365$     &  $40.8$  &    $0.6$    & I   & $8$   & $+349$   & $4.5$ \\
$f$  &  $-280$     &  $31.3$  &    $0.8$    & II  & $2$   & $-275$   & $1.8$ \\
$g$  &  $+203$     &  $22.7$  &    $1.1$    & II  & $3$   & $+198$   & $2.5$ \\
$h$  &  $-148.5$   &  $16.6$  &    $1.5$    & II  & $4$   & $-155.2$ & $4.5$ \\
$i$  &  $+147.3$   &  $16.5$  &    $1.5$    & II  & $4$   & $+155.2$ & $5.1$ \\
$j$  &  $+128.1$   &  $14.3$  &    $1.7$    & II  & $5$   & $+127.6$ & $0.4$ \\
$k$  &  $-103.6$   &  $11.6$  &    $2.2$    & II  & $6$   & $-108.4$ & $4.6$ \\
$l$  &  $+99.9$    &  $11.2$  &    $2.2$    & II  & $7$   & $+94.5$  & $5.4$ \\
$m$  &  $-79.6$    &  $8.9$   &    $2.9$    & II  & $8$   & $-83.9$  & $5.4$ \\
$n$  &  $+72.8$    &  $8.1$   &    $3.1$    & II  & $9$   & $+75.4$  & $3.6$ \\
$o$  &  $-67.0$    &  $7.5$   &    $3.4$    & II  & $10$  & $-68.6$  & $2.3$ \\
$p$  &  $+64.1$    &  $7.2$   &    $3.6$    & III & $6$   & $+64.0$  & $0.1$ \\
$q$  &  $+60.4$    &  $6.7$   &    $3.9$    & II  & $11$  & $+62.9$  & $4.1$ \\
$r$  &  $-55.5$    &  $6.2$   &    $4.2$    & III & $7$   & $-55.9$  & $0.7$ \\
$s$  &  $+51.2$    &  $5.7$   &    $4.6$    & III & $8$   & $+49.7$  & $3.0$ \\
$t$  &  $-44.0$    &  $4.9$   &    $5.5$    & III & $9$   & $-44.8$  & $1.8$ \\
$u$  &  $+41.5$    &  $4.6$   &    $5.9$    & III & $10$  & $+40.8$  & $1.6$ \\
$v$  &  $-37.7$    &  $4.2$   &    $6.6$    & III & $11$  & $-37.5$  & $0.5$ \\
$w$  &  $+34.3$    &  $3.8$   &    $7.3$    & III & $12$  & $+34.7$  & $1.1$ \\
$x$  &  $+29.3$    &  $3.3$   &    $8.8$    & III & $14$  & $+30.2$  & $3.0$ \\
$y$  &  $-28.7$    &  $3.2$   &    $9.0$    & III & $15$  & $-28.3$  & $1.3$ \\
$A$  &  $+19.4$    &  $2.2$   &    $14$     & IV  & ?     & ?        & ?     \\
$B$  &  $-18.0$    &  $2.0$   &    $15$     & IV  & ?     & ?        & ?     \\\hline 
\end{tabular}
\label{tab:sh}
\caption[Shells in NGC\,3923  and their identification found in \citet{bil13}]{Shells in NGC\,3923  and their identification found in \citet{bil13} (data from \citealp{bil13} and \citealp{bil14}). $\lambda$\,--\,Des\-ignation of the shell. $R$\,--\,Observed radius of the shell  (data taken from \citealp{prieur88} and \citealp{sikkema07}). The plus sign means that the shell is situated on the northern side of the galaxy and minus on the southern; $a$\,--\,Gravitational acceleration at the edge of the shell in the potential used in \citet{bil13} expressed in the units of the MOND acceleration constant. $G$\,--\,Identified generation number of the shell. $n$\,--\,Identified serial number of the shell.  $r$\,--\,Modeled radius of the shell. $\Delta$\,--\,Relative difference between the observed and the modeled shell radius.}
\end{table}
\clearpage

To obtain the MOND acceleration filed within the galaxy, we started from its archival near-infrared image taken by the Spitzer Space Telescope in the 3.6\,$\mu$m band. We chose this band because it is little affected by extinction and because the stellar mass-to-light ratio varies only mildly in it. We fitted the galaxy by a~smooth analytic profile. As follows from \citet{angus13}, it is necessary to model the mass distribution accurately when testing MOND. For this reason we fitted the surface-brightness profile by a~sum of two \sers profiles. Next, we assumed that the galaxy is a~prolate ellipsoid with its major axis oriented perpendicularly to the line of sight. This is motivated by the facts that 1) the galaxy shows minor-axis rotation \citep{minrot}, 2) this type morphology of a~shell system is more probable for a~prolate galaxy (see \fig{prolobl} and \citealp{DC86}), and 3) the shells would not be observable from a~substantially different viewing angle. We used the distance of the galaxy from Earth of 23\,Mpc -- the median of all the published results of the direct methods. To convert the surface brightness to the surface density, we needed to know the mass-to-light ratio. We derived the $J$ and $K_\mathrm{S}$ magnitudes of the galaxy from its images taken in the 2MASS survey and used Eq.~(4) of \citet{for12} to obtain the mass-to-light ratio in the 3.6\,$\mu$m band. We checked that the contribution of coronal interstellar gas is negligible in this galaxy. We assumed that the mass-to-light ratio is constant through the galaxy. Then we were able to deproject the surface density to the volume density and to calculate the Newtonian gravitational field produced by the baryonic matter. We calculated the MONDian acceleration field in the galaxy using the algebraic relation \equ{algrel} with the simple interpolating function \equ{musimp}. Then we calculated the expected shell evolution in this potential from Eqs.~\ref{eq:hqform}, \ref{eq:tate} and \ref{eq:rn}. By doing so, we assume that MOND is a~theory of modified gravity or a~correlation between dark and luminous matter. If MOND is a~theory of modified inertia instead, then the immediate acceleration of a~body generally depends on the history of its motion, i.e. a~gravitational potential satisfying $- \nabla \phi = \ddot{\mathbf{r}}$ cannot be constructed.

It came out that the 25 largest observed shells could be indeed divided into three groups, corresponding to three generations, so that each of them was reproduced by the model at a~certain time, see \fig{shcomp}. The identified serial and generation numbers are noted in \tab{sh}. There are two missing shells in the first generation and one in the third generation.

Apart from the required properties described in \sect{shid}, this shell identification has several extra desirable features. 1) Given that the gravitational potential had no free parameters, the coincidence of the modeled and the observed radii is surprisingly good ($\leq 5.4$\% deviation). 2) Almost all shells from the $N$-th generation lie on larger radii than the shells from $N+1$-st generation. This is in accordance with the effect of dynamical friction. 3) We needed to suppose the existence of only three missing shells. At the expected position of one of them, there is a~dust cloud visible in Fig. D.1 of \citet{sikkema07}.

The most serious downside of this identification was that the two innermost shells could not be reproduced by the model. Maybe some of the assumptions of our model could be broken for them: the merger could be slightly non-radial or the secondary decayed sooner than it reached the center of the primary galaxy for the last time. The other downside was that one of the shells, denoted as $i$ in \tab{sh},  had to be interpreted as the shell $h$ encircling the whole galaxy. The two have almost the same radius. But why there is only one shell like this? Some of our unpublished simulations showed that the shells encircling the whole galaxy form easily, but if the shell $n$ encircles the galaxy, then also the shells with the higher serial numbers encircle it as well. Maybe this has something to do with the focusing effect of elliptical potentials (see \sect{sim}) -- our simulations included only spherical potentials. Nevertheless, this problem should be addressed in future.

Except for these unclear aspects, the paper B13 confirmed the consistency of the shell distribution in NGC\,3923 with MOND. We arrived to the opposite conclusion from \citet{hqmond} because of the better observational data and the better understanding of the shell formation process (multiple-generation formation).

\subsection{MOND prediction of a~new shell in NGC\texorpdfstring{\,3923}{Lg}}\label{sec:bil14}
The shell identification found in B13 implied the existence and positions of as yet unobserved shells. The most interesting  one was the shell number one of the first generation. It was expected to lie approximately 200\,kpc from the center of NGC\,3923. If discovered, it would be the biggest shell ever observed and a~support for MOND. It would confirm MOND down to acceleration of $a_0/10$, an exceptionally low value for an elliptical galaxy. 

But the shell identification from B13 was found only for one fixed gravitational potential and it was found ``by eye''.  To make a~strong prediction, we needed to explore all gravitational potentials permitted by observations and to find all acceptable shell identifications for them. This would be extremely time consuming if the shell identifications would be searched manually. Moreover, we could easily miss some acceptable shell identifications. For this reason, we created a~program for the automatic shell identification.

\begin{figure}
	\centering\includegraphics[width=0.5\textwidth]{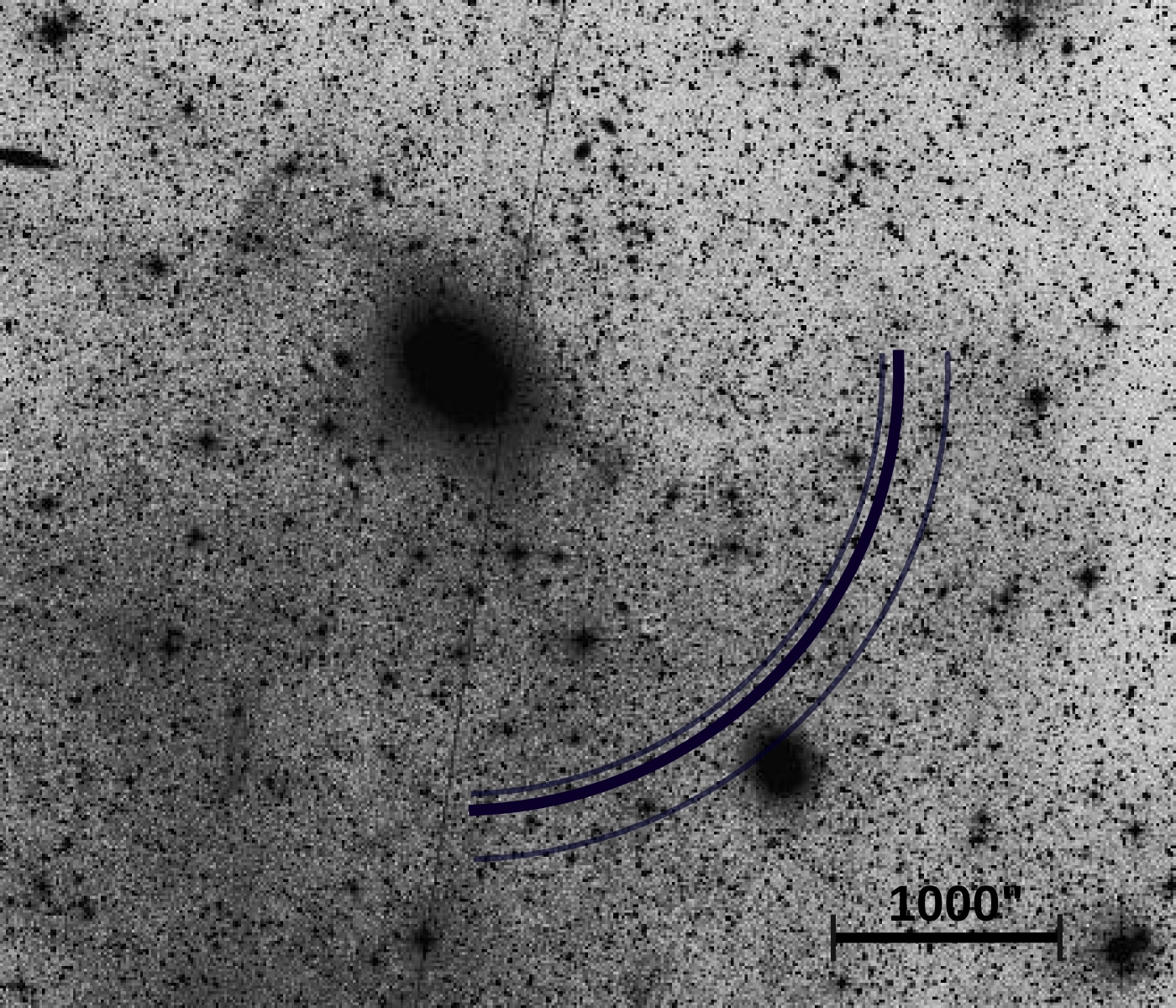}
\caption[Predicted position of the new shell in NGC\,3923 by \citet{bil14}]{Predicted position of the new shell in NGC\,3923 by \citet{bil14}. }
\label{fig:nsh}
\end{figure}

The program tests the compatibility of the given gravitational potential with the observed shell distribution in a~Type~I shell galaxy. The program takes as the input: 1) The tested potential; 2) The radii of the observed shells with the sign added according to the galaxy side; 3) The acceptable deviation between the observed and modeled radius for each observed shell; 4) The number of generations in the system; 5) The maximal allowed number of missing shells in each individual generation. The program returns all acceptable shell identifications for the tested potential, i.e. those meeting the criteria described in \sect{shid}.

In the paper \citet{bil14} (B14 hereafter, see the paper attached in \app{bil14} for details) we used the program to verify our preliminary expectation about the new giant shell. We varied many free parameters of the potential: The distance of \ngc{} from Earth, its mass-to-light ratio, the value of the MOND acceleration constant $a_0$, the choice of the MOND interpolating function, and the characteristic radius of the galaxy. We also tried to treat the shells $h$ and $i$ from \tab{sh} either as two individual shells or as a~single shell encircling the galaxy. We considered only the 25 outermost shells. In total, we searched shell identifications for several hundreds of gravitational potentials. 

All the acceptable identifications we found had a~common property: The four outermost shells ($a$, $b$, $c$ and $d$ in \tab{sh}) had to have the serial numbers 2, 3, 4 and 5 and they all had to come from the first generation. This meant that the shell number 1 had to exist. Its radius came out to be 1930-2090\arcsec{}, depending on the choice of the free parameters (see \fig{nsh}). The most probable value (i.e. that for the choice of the parameters by B13) was 1900\arcsec{}.

The code was successfully tested using a~Newtonian simulation. It was able to recover the serial and generation numbers correctly.  

We also tested the predictive ability of the method. We excluded the outermost currently known shell ($a$ in \tab{sh}) and tried to reconstruct its radius from the positions of the remaining shells by the same way as before. The result was was almost precise. The calculated value was 1160\arcsec{} while the correct value is 1170\arcsec{} (1\% deviation).

Since the expected linear size of the predicted shell was about 2 times larger than that of the shell $a$, its surface brightness should be roughly 4 times lower, supposing they are made of the same number of stars. Hence, the expected surface brightness of the predicted shell was about 28\,mag\,arcsec$^{-2}$ in the B~band. This is low enough for the shell to be easily missed by the previous observations. The deepest images of this region were taken using photographic plates in the 80s (e.g., \citealp{MC83}). These images reached the surface brightness limit of around 26.5\,mag\,arcsec$^{-2}$ in B. If the shell  had been found, it would clearly be a~supporting argument for MOND. If no new shell had been found in the corresponding region, it would have no implication for the validity of MOND since the surface brightness of the predicted shell could be below our detection limit. But if the shell had had a~substantially different radius, or two or more new shells had been found in the corresponding region, it would be an argument against MOND.

This prediction constituted a~rare test of MOND in an elliptical galaxy down to the gravitational acceleration of around $a_0/10$. There are only two individual ellipticals where MOND could be tested down to such a~low acceleration \citep{milg12}. If the shell had been discovered it would have been the first discovery of an object on the basis of MOND, having a~lot of common with the discovery of Neptune in the early stages of Newtonian dynamics.

\subsection{The search for the shell predicted by MOND}\label{sec:bil15b}
We got observing time at the MegaCam camera mounted on the 3.6\,m Canada-France-Hawaii Telescope (CFHT) to search for the predicted shell (PI Michal B{\'i}lek\footnote{{B{\' i}lek}, M., {Jungwiert}, B. and {J{\' i}lkov{\' a}}, L.: Program 14BO12 -- ``Ultra-deep wide-field imaging of a~shell elliptical NGC\,3923 with
MegaCam: looking for a~new shell at $\sim$220\,kpc and constraining gravitational potential'', proposed via OPTICON in 2014}). This instrument had got famous previously for the ultra-deep images of Virgo Cluster (New Generation Virgo Survey, \citealp{ngvs}) or the discoveries of faint tidal features in elliptical galaxies (the project \atlas, \citealp{atlas3d1}). These images reached the surface brightness limit of 29\,mag\,arcsec$^{-2}$ in the $g'$ band. The camera has the field-of-view of $1\times 1$ degree.

\begin{figure}[t]
	\centering\includegraphics[width=0.5\textwidth]{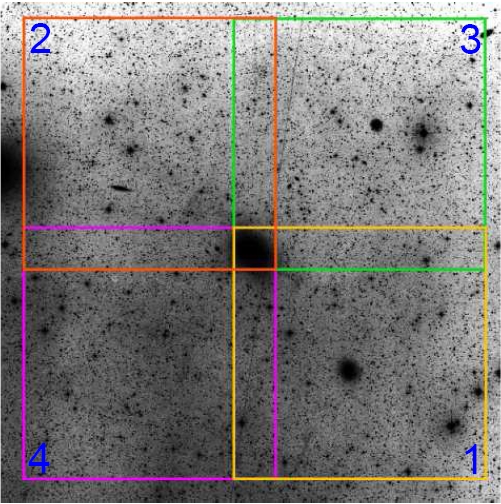}
\caption[Mosaic used when imaging the galaxy NGC\,3923 by \citet{bil15b}]{Arrangement of the mosaic used to image the galaxy NGC\,3923 by \citet{bil15b}. The underlying photograph comes from the DSS survey. }
\label{fig:tiles}
\end{figure}

We presented our observations in \citet{bil15b} (see the paper attached in \app{bil15b} for details). The galaxy was covered by a~four-tile mosaic centered on the galaxy, see \fig{tiles}. We chose the $g'$ band, because the previous works \citep{duc11} proved that it is the best for reaching the lowest surface-brightness limit. Having the total exposure time of 273\,min, our image reached the limit of 29\,mag\,arcsec$^{-2}$. 

To complete our study of the tidal features in the galaxy, we also reanalyzed an archival image taken by Hubble Space Telescope (HST) presented by \citet{sikkema07}. Because of its higher angular resolution, it revealed the shells in the center of NGC\,3923 better than the MegaCam image. This image was taken in the F814W and F606W bands. Its total exposure time was 35\,min. 

\begin{figure}
	\centering\includegraphics[width=1\textwidth]{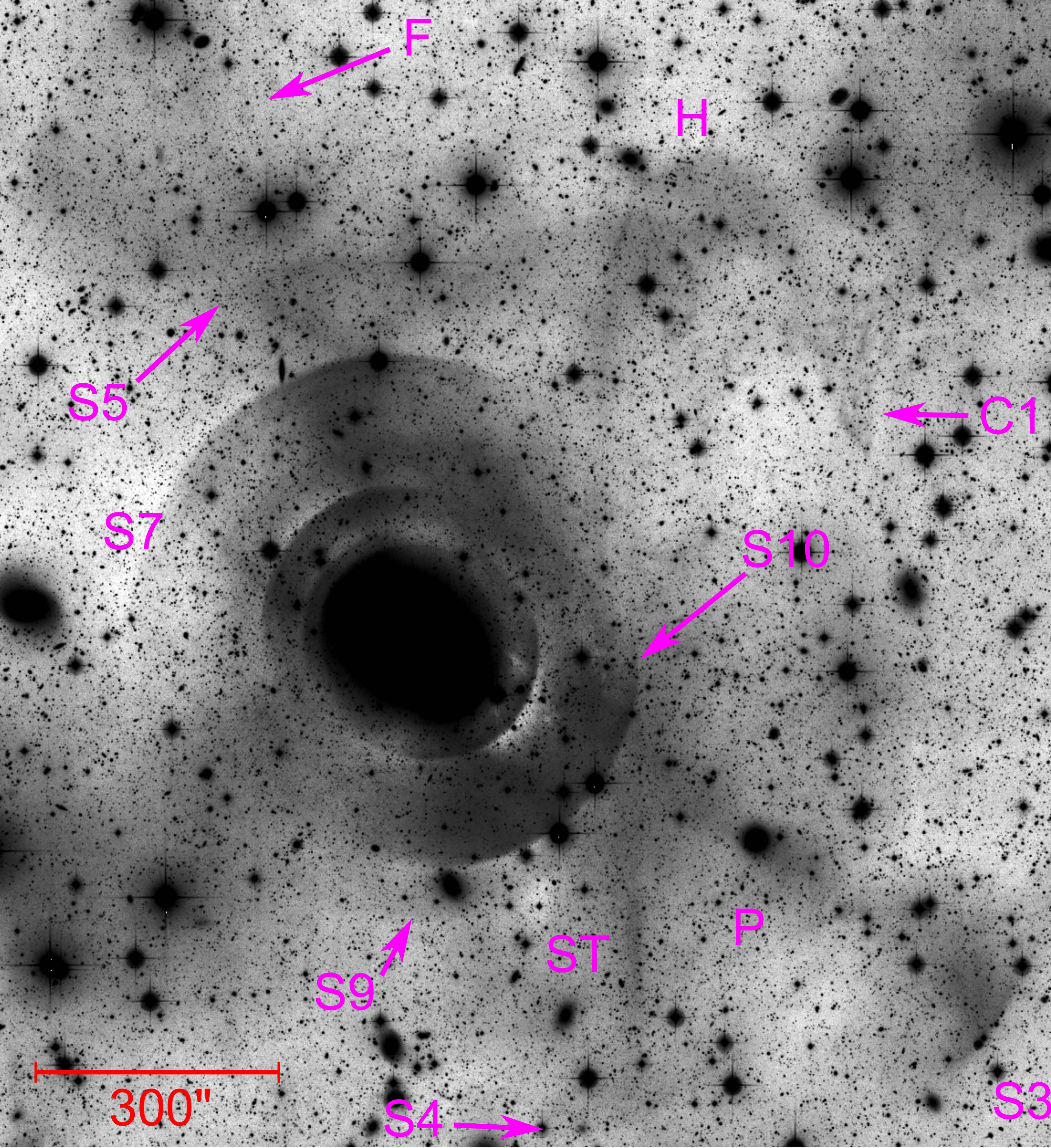}
\caption[Galaxy NGC\,3923 imaged by \citet{bil15b}]{Galaxy NGC\,3923 imaged by \citet{bil15b} using the MegaCam camera at 3.6\,m Canada-France-Hawaii Telescope. Shells are marked by S$n$. The progenitor of the shell S3 is marked by P. The signs F, H and ST are other (non-shell) tidal features. There was also a~Galactic cirrus C1 captured (a dust cloud in our own Galaxy). }
\label{fig:tidal}
\end{figure}

We found 42 shells in the galaxy in total. This is substantially more than thought before (the compilation by \citealp{bil13} contained 27 shells). Two of the shells used for the prediction of the new shell by B14 do not exist at all (the shells $c$ and $d$ in \tab{sh}). We found the probable progenitor of at least one of the shells (the shell $b$ in \tab{sh}), see \fig{tidal}. The shell predicted by B14 was not detected.

Since the ratio of the projected radii of the progenitor and the innermost clearly detected shell is 26, the shell system was likely created by two or more progenitors. This follows from the findings from simulations summarized in \citet{DC87} (see \sect{sim}): It was found that the multiple-generation shell formation is required to produce  shell systems with the radial range higher than about 6 in the minor merger model. The largest shells form first, while the inner shells form later, when the secondary loses enough orbital energy by dynamical friction. But even if the observed progenitor is in the apocenter now, it has enough energy to create some of the largest shells in NGC\,3923 in its next pericentric passage. Thus, it seems that the shells had to be created by more than one progenitor. This is supported by the fact that the number of shells in \ngc{} is the highest of all galaxies. The papers B13 and B14 were based on the assumption that the shell system originates from one secondary.

Thus, it is not surprising that the predicted shell was not found. The prediction was based on incomplete and erroneous data and incorrect assumptions. For this reason, the shell radial distribution in NGC\,3923 has no implication for the validity of MOND at this stage. A~new analysis will have to be done. 

But it is interesting that we detected so many shells, even if they come from two secondaries. Perhaps this could be also used for distinguishing between the MOND theories and the DMF. As we know, only a~few shells can be made per generation (see \sect{sim}). Perhaps the high number of shells indicates that the secondaries had to make a~lot oscillations to create that many shells. This could then mean that the dynamical friction was very low and therefore MOND is preferred. But to make a~stronger conclusion here, we would need to perform a~number of self-consistent simulations of shell galaxy formation both in MOND and the DMF.

\subsection{Velocities derivable from shell spectra in MOND}\label{sec:bil15a}
Not only the shell radii can be used for constraining the gravitational potential of shell galaxies. The line profiles in shell spectra offer another way (see the papers \citealp{bil15a} and also \citealp{bilcjp} in \app{bil15a} and \app{bil14b}, respectively, for details). The idea originally comes from \citet{merrifieldkuijken}. \citet{ebrova12} deduced that the spectral line profiles have four peaks (the quadruple-peaked profile). The separations of the peaks are given by the circular velocity at the edge of the shell and the expansion velocity of the shell itself. The corresponding formulas can be found in \citet{ebrova12}. These formulas are valid provided that the potential of the galaxy is spherical, the merger was exactly radial, and the secondary decayed in the primary center. These velocities are quantities tightly connected with the gravitational potential $\phi$ of the host galaxy. The circular velocity at the radius $r$ is
\begin{equation}
	v_\mathrm{c}(r) = \sqrt{r\frac{\d \phi(r)}{\d r}}.
\end{equation}

As derived in \sect{shrad}, the expansion velocity of the $n$-th shell at the radius $r$ is
\begin{equation}
 v_{\mathrm{e},n}(r) \approx v_{\mathrm{A},n}(r) =  \frac{1}{\left(n+1/2\right)\frac{\d P(r)}{\d r}},
\label{eq:ve}
\end{equation}
where the period of oscillation $P(r)$ is given by
\begin{equation}
	P(r) = \sqrt{2}\int_0^r \left[\phi(r)-\phi(x)\right]^{-1/2}\d x.
	\label{eq:pr2}
\end{equation}
The expansion velocity thus depends on the whole profile of the potential from the center of the galaxy to the edge of the shell and on the serial number of the shell.

The observations of the shell spectral-line profiles are probably beyond the capabilities of the existing instruments. The main problem is the low surface brightness of the shells. They could however become observable in near future with the instruments being planned.

In the paper \citet{bil15a} (B15c hereafter, see also \citealp{bilcjp}), we were interested in the asymptotic behavior of these two velocities in MOND at large radii. The expression for the asymptotic circular velocity $V_\mathrm{c}$ is obviously the same as for the disk galaxies:
\begin{equation}
	V_\mathrm{c} = \sqrt[4]{GMa_0},
	\label{eq:vca}
\end{equation}
where $M$ is the total baryonic mass of the galaxy. As we know (see \sect{impl}), this formula expresses the baryonic Tully-Fisher relation. The tests of this relation are rare in the ellipticals because these galaxies mostly miss rotating tracers at large radii. When shell spectra become available, they will help us to rectify this insufficiency. It is expected by MOND theories that the baryonic Tully-Fisher relation holds true for the disk and the elliptical galaxies equally. In the DMF there is no obvious reason for this: The circular velocity is dominated by the distribution of DM at large radii. And since the mass-assembly histories of the disk and elliptical galaxies are expected to be different, they are not expected to meet the same baryonic Tully-Fisher relation.

\begin{figure}
	\centering\includegraphics[width=0.6\textwidth]{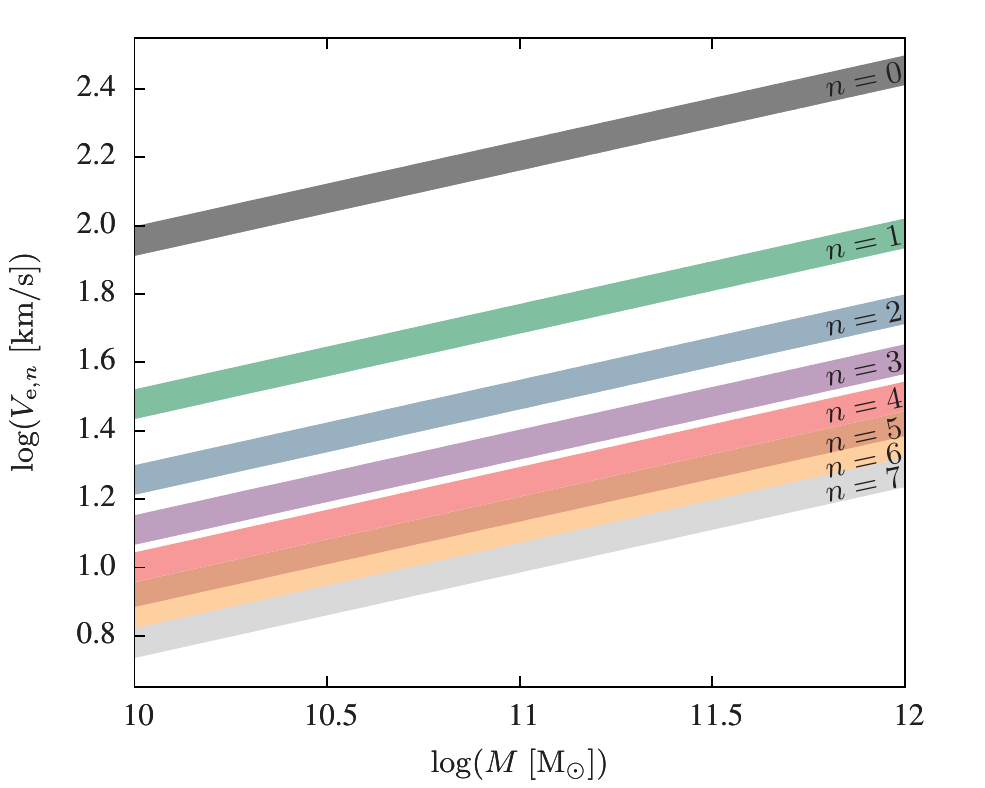}
\caption[Analogy of the baryonic Tully-Fisher relation for shell expansion velocities]{Analogy of the baryonic Tully-Fisher relation for shell expansion velocities. Each branch corresponds to a~certain shell serial number. The variable $M$ stands for the total baryonic mass of the galaxy and $V_{\mathrm{e},n}$ for the asymptotic expansion velocity of the $n$-th shell. The width of the bands is $\pm 10\%$ of the value given by \equ{vea}. }
\label{fig:tfshell}
\end{figure}

Let us derive the expression for the asymptotic expansion velocity of a shell in MOND. The MOND gravitational potential behaves asymptotically like $\phi_\mathrm{L} = \sqrt{GMa_0}\log(r)+\phi_0$, where $\phi_0$ is a normalization constant (again, we assume that a MOND potential can be constructed). Employing \equ{pr2}, the oscillation period is
\begin{equation}
	P_\mathrm{L}(r) = \frac{\sqrt{2\pi}}{\sqrt[4]{GMa_0}}\,r
	\label{eq:prlog}
\end{equation}
in this logarithmic potential. The real MOND potential differs from $\phi_\mathrm{L}$ at low radii. But recall that the oscillation period $P(r)$ is twice the free-fall time from the radius $r$. Potential wells of isolated objects are infinitely deep in MOND. If a~particle falls from a~sufficiently large radius, it can reach an arbitrarily high velocity at a~given distance from the center and move across the inner problematic region in an arbitrarily short time. For this reason, the real oscillation period in MOND is also given by \equ{prlog} for large radii.

Substituting \equ{prlog} into \equ{ve} we finally get the shell expansion velocity at very large radii in MOND
\begin{equation}
	V_{\mathrm{e},n} = \frac{\sqrt[4]{GMa_0}}{(n+1/2)\sqrt{2\pi}} =  \frac{V_\mathrm{c}}{(n+1/2)\sqrt{2\pi}}.
	\label{eq:vea}
\end{equation}
The first equality in \equ{vea} expresses an analogy of the baryonic Tully-Fisher relation for the shell expansion velocities. This relation has several branches, see \fig{tfshell}. Each of them corresponds to a certain shell serial number.

If the ratio of the circular and expansion velocities were measured for a~lot of sufficiently large shells, the histogram of this ratio would form a~series of equidistant peaks. The separation of the neighboring maxima would be $\sqrt{2\pi}$.

The results stated up to this point are precise only for very large shells. In the paper B15c, we were also interested what ``very large'' means. Namely, above which radius the circular or expansion velocity differs by less than 10\% from its asymptotic value. The answer depends, of course, on the mass distribution of the galaxy. Most shells are in early-type galaxies and the baryonic matter in most early-type galaxies follows well the \sers law \citep{sersic}. 

The surface-brightness profile of a~\sers sphere is defined as
\begin{equation}
	\Sigma(r) = \Sigma_\mathrm{e}\exp{\left\{-b_\nu\left[\left(\frac{r}{r_\mathrm{e}}\right)^{1/\nu}-1\right]\right\}}.
	\label{eq:sers}
\end{equation}
The free parameters are the \sers index $\nu$, the effective radius $r_\mathrm{e}$ and the surface brightness at the effective radius $\Sigma_\mathrm{e}$. The variable $b_\nu$ is a~function of $\nu$. It is chosen so that the circle of the radius $r_\mathrm{e}$ contains one half of the total flux. Given the mass-to-light ratio, we can obtain the volume density profile of a~\sers sphere by deprojection. In the paper B15c, we used the approximate analytic deprojection of the \sers profile published by \citet{sersdeproj}.

\begin{figure}
	\centering\includegraphics[width=1\textwidth]{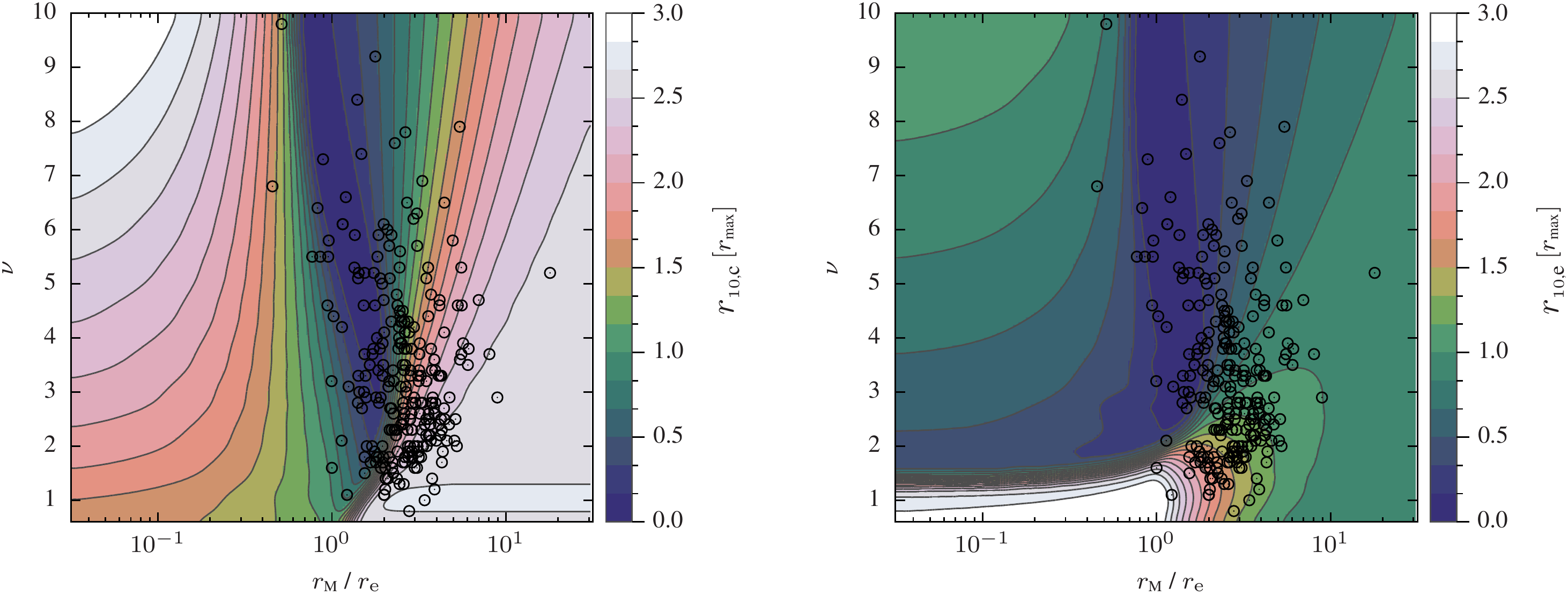}
\caption[Radii of approaching the asymptotic velocity]{Left -- The radius $r_{10, \mathrm{c}}$ above which the circular velocity differers by less than 10\% from its asymptotic value given by \equ{vca}. Right -- The radius $r_{10, \mathrm{e}}$ above which the shell expansion velocity differers by less than 10\% from its asymptotic value given by \equ{vea}. These radii are expressed in the units or $r_\mathrm{max}$, which is the maximum of the half-light radius $r_\mathrm{e}$ and the MOND radius $r_\mathrm{M}$. The maps were calculated for a~grid of \sers spheres characterized by $r_\mathrm{e}$ and the \sers index $\nu$. The circles show the distribution of the real galaxies from the \atlas sample in these maps. From \citet{bil15a}. }
\label{fig:z10}
\end{figure}

Let us denote by $r_{10,\mathrm{e}}$ the radius above which the shell expansion velocity differs by less than 10\% from its asymptotic value, and similarly $r_{10,\mathrm{c}}$ for the circular velocity. The symbol $r_{10}$ stands for $r_{10,\mathrm{c}}$ or $r_{10,\mathrm{e}}$ whenever they can be used interchangeably.  From the dimensional grounds, we have
\begin{equation}
	r_{10} = r_\mathrm{e}\,y_{10}\left(\theta,\nu\right).
	\label{eq:y10}
\end{equation}
The function $y_{10}(\cdot, \cdot)$ is to be recovered numerically. The parameter $\theta$ is defined as $r_\mathrm{M}/r_\mathrm{e}$, where $r_\mathrm{M} = \sqrt{GM/a_0}$ is the MOND radius.

We defined a~grid of the parameters $\theta$ and $\nu$ and calculated $r_{10}$ at its every note numerically (see \fig{z10}). We also compared the grid with the real distribution of $\theta$ and $\nu$ for a~complete sample of nearby galaxies \atlas  \citep{atlas3d1}.

Let us define $r_\mathrm{max} = \max\left(r_\mathrm{e}, r_\mathrm{M}\right)$. It came out that $r_{10}$ is between 0 and 3 $r_\mathrm{max}$. The radius $r_{10,\mathrm{c}}$ is greater than $r_{10,\mathrm{e}}$ for most galaxies. 

If we test the baryonic Tully-Fisher relation or its analogy for shell expansion velocities, we should use only the shells greater than $r_{10, c}$ or $r_{10, e}$ not to introduce systematic errors. The latter statement holds true for any type of measurement of the circular velocity, not only those based on the shell spectra.

It is surprising that $r_{10}$ came out zero for many galaxies, i.e. the rotation or expansion curve is almost flat from the galaxy center. The mass distribution in these galaxies just compensates the transition from the Newtonian to the deep-MOND regime. This is the case of the galaxies having $\theta \approx 1$ and $\nu \gtrsim 2$. Such galaxies really exist, one example is NGC\,3923.

\section{Summary}
MOND is an observationally deduced rule for predicting the acceleration of bodies from the distribution of the observable matter (\sect{mond}).  This rule works surprisingly well in stellar systems from star clusters to all types of galaxies. The explanation of such a~tight and universal correlation is unclear for Newtonian dynamics with dark matter. The simplest explanation is that it is a~consequence of a~new law of physics -- a~theory of MOND. MOND theories are based on postulating the breakdown of the Newtonian dynamics for accelerations lower than  $a_0 = 1.2\times10^{-10}$\,m\,s$^{-2}$ where the space-time scaling symmetry emerges. They include modified gravity and modified inertia theories. Nevertheless, a~lot of implications follow from the basic tenets alone (\sect{impl}) regardless of the theory. Other findings about MOND follow from galaxy simulations (\sect{mondsim}). However, MOND also deals with theoretical and observational shortcomings (\sect{mondprob}).  

Very little tests of MOND down to low accelerations were done in elliptical galaxies. This work summarized the methods and ideas the author developed to test MOND in the elliptical galaxies containing stellar shells. The methods can also be used for constraining the dark matter distribution supposing the Newtonian dynamics. These shells are made of stars on nearly radial orbits. This is very interesting from the point of view of MOND because MOND was originally inspired by the disk galaxies where the kinematic tracers move on circular orbits. The investigation of shell galaxies could therefore eliminate or prefer some of the MOND modified inertia theories. Also the difference in the effectivity of dynamical friction in some of the MOND theories and Newtonian dynamics probably has observable impact on the shell system appearance, which can be investigated by self-consistent simulations (\sect{mondsh}). 

We listed the observational properties of shell galaxies (\sect{obs}) and reviewed shortly the main formation scenarios including the advantages and disadvantages of each of them (\sect{form}). In the rest of the work, we assumed the phase-wrapping minor merger scenario, which corresponds to the observing data best. We summarized the findings from the simulations of shell formation by this mechanism (\sect{sim}). One of them indicates that for a~given potential, the evolution of shell radii in the axially symmetric shell systems can be modeled analytically. We described various methods of such modeling from the simplest approximations to the precise formulas (\sect{shrad}). Some of them have never been published so far.

The rest of the thesis summarizes the author's published original results. One of the main achievements is the method of shell identification which allows testing the consistency of a~given potential with the observed shell radial distribution in the galaxies possessing axially symmetric shell systems (\sect{shid}). We used it to verify the consistency of the shell distribution in the elliptical galaxy NGC\,3923 with MOND down to a~very low acceleration (\sect{bil13}). Moreover, the method could be used to predict the existence of a~new as yet undiscovered shell in NGC\,3923 (\sect{bil14}). It could escape the previous observations easily because the deepest images of the corresponding region came from the 80s. If discovered, it would be the biggest shell ever observed and the first discovery of an object predicted by MOND. We get observing time at 3.6\,m Canada-France-Hawaii Telescope and obtained ultra-deep images of the galaxy reaching the surface-brightness limit of 29\,mag\,arcsec$^{-2}$ to search for the predicted shell (\sect{bil15b}). The shell was not there but we found that our previous works were based on poor data. The analysis will have to be redone. Nevertheless, we were able to conclude that the shell system in NGC\,3923 was created from two progenitors at least and one of them is still observable. The high total number of shells detected (42) is very interesting by itself and may come out to be important for discriminating between the MOND theories and the Newtonian dynamics with dark matter.

Line profiles in shell spectra offer another opportunity to test MOND in the ellipticals. They will not be however available until the next generation of instruments comes. The spectra bring information about the expansion velocity of the shell and the circular velocity at the shell edge. We derived that in MOND, these velocities are expected to be determined by the total baryonic mass of the galaxy for very large shells (\sect{bil15a}).

\newpage

\setlength{\bibsep}{0pt plus 0.3ex}
\bibliographystyle{klunamed}
\bibliography{citace}

\newpage

\listoffigures
\listoftables

\newpage
\appendix
\openright
\section{Author's publications}
\textit{(Copies of the papers are attached in the original version of the thesis)}
\subsection{B{\' i}lek et al., 2013} \label{ap:bil13}
The paper \textbf{{B{\'{\i}}lek}}, \textbf{M.}, {Jungwiert}, B., {J{\'{\i}}lkov{\'a}}, L., 
	{Ebrov{\'a}}, I., {Barto{\v s}kov{\'a}}, K., \& {K{\v r}{\'{\i}}{\v z}ek}, M.: 2013,  `Testing
MOND gravity in the shell galaxy NGC 3923', \aap, 559, A110 \\ \url{http://adsabs.harvard.edu/abs/2013A\%26A...559A.110B}

\openright
\subsection{B{\' i}lek et al., 2014} \label{ap:bil14}
The paper \textbf{B{\'{\i}}lek}, \textbf{M.}, Barto{\v s}kov{\'a}, K., Ebrov{\'a}, I., \& Jungwiert, B.: 2014, `MOND prediction of a new
giant shell in the elliptical galaxy NGC 3923'. \aap, 566, A151 \\ \url{http://adsabs.harvard.edu/abs/2014A\%26A...566A.151B}

\openright
\subsection{B{\' i}lek et al., 2015a} \label{ap:bil15b}
The paper  {\textbf{B{\'{\i}}lek}},\textbf{ M.}, {Cuillandre}, J.-C., {Gwyn}, S.,	{Ebrov{\'a}}, I., {Barto{\v s}kov{\'a}}, K., {Jungwiert}, B., \&	{J{\'{\i}}lkov{\'a}}, L.: 2015, `Deep imaging of the shell elliptical galaxy NGC3923 with MegaCam'.  arXiv:1505.07146, submitted to \aap.
\\ \url{http://adsabs.harvard.edu/abs/2015arXiv150507146B}

\openright
\subsection{B{\' i}lek et al., 2015c} \label{ap:bil15a}
The paper   \textbf{B{\'{\i}}lek}, \textbf{M.}, Jungwiert, B., Ebrov{\'a}, I., \& Barto{\v s}kov{\'a}, K.,  2015: `MOND implications for
spectral line profiles of shell galaxies: shell formation history and mass-velocity scaling relations'. \aap, 575, A29 
\\ \url{http://adsabs.harvard.edu/abs/2015A\%26A...575A..29B}

\openright
\subsection{B{\' i}lek et al., 2015b} \label{ap:bil14b}
The paper  \textbf{B{\'{\i}}lek}, \textbf{M.}, Ebrov{\'a}, I., Jungwiert, B., J{\'{\i}}lkov{\'a}, L., \& Barto{\v s}kov{\'a}, K.: 2015, `Shell galaxies as laboratories for testing MOND'. Canadian Journal of Physics, 93, 203
\\ \url{http://adsabs.harvard.edu/abs/2015CaJPh..93..203B}

\openright
\subsection{Ebrov{\' a} et al., 2012} \label{ap:ebrova12}
The paper {Ebrov{\'a}}, I., {J{\'{\i}}lkov{\'a}}, L., {Jungwiert}, B., 
	{K{\v r}{\'{\i}}{\v z}ek}, M., \textbf{{B{\'{\i}}lek}}, \textbf{M.}, {Barto{\v s}kov{\'a}}, K., 
	{Skalick{\'a}}, T., \& {Stoklasov{\'a}}, I.: 2012, `Quadruple-peaked spectral line profiles as a tool to constrain gravitational potential of shell galaxies. \aap, 545, A33 
\\ \url{http://adsabs.harvard.edu/abs/2012A\%26A...545A..33E}

\openright
\end{document}